\documentclass[a4paper,11pt,preprintnumbers]{article}
\pdfoutput=1

\usepackage{multirow}
\usepackage{amssymb,mathrsfs,amsmath}
\usepackage{a4wide}
\usepackage{color,xcolor}
\usepackage{slashed,soul}
\usepackage{graphicx}
\usepackage{amsfonts}
\usepackage{cancel}
\usepackage{lscape}
\usepackage{float}
\def\linkcolor{cyan!70!black}

\usepackage[
colorlinks=true
,urlcolor=\linkcolor
,anchorcolor=\linkcolor
,citecolor=\linkcolor
,filecolor=\linkcolor
,linkcolor=\linkcolor
,menucolor=\linkcolor
,linktocpage=true
,pdfproducer=medialab
,pdfa=true
]{hyperref}
\usepackage{amsthm}
\usepackage{booktabs}
\usepackage{array}
\usepackage{rotating}
\usepackage[numbers, sort&compress]{natbib}
\usepackage{multirow}
\usepackage{float}
\usepackage[utf8]{inputenc}
\usepackage[T1]{fontenc}
\usepackage{appendix}
\usepackage{epsfig}
\usepackage{orcidlink}
\usepackage{comment}
\usepackage{tabularx}

\usepackage{ulem}

\newcommand{\beq}{\begin{equation}} 
\newcommand{\eeq}{\end{equation}} 
\newcommand{\ba}{\begin{array}}  
\newcommand{\ea}{\end{array}} 
\newcommand{\bea}{\begin{eqnarray}}  
\newcommand{\eea}{\end{eqnarray} }  
\newcommand{\bal}{\begin{align}}
\newcommand{\eal}{\end{align}}   
\newcommand{\bi}{\begin{itemize}}  
\newcommand{\ei}{\end{itemize}}  
\newcommand{\ben}{\begin{enumerate}}  
\newcommand{\een}{\end{enumerate}}  
\newcommand{\bc}{\begin{center}}
\newcommand{\ec}{\end{center}} 
\newcommand{\bt}{\begin{table}}
\newcommand{\et}{\end{table}}  
\newcommand{\btb}{\begin{tabular}}
\newcommand{\etb}{\end{tabular}}

\newcommand{\tzpdf}{\texorpdfstring{$\theta_{23}$ }{tz}}

\newcommand{\chipdf}{\texorpdfstring{$\chi^2$ }{chisq}}
\newcommand{\dcppdf}{\texorpdfstring{$\delta_{CP}$ }{deltaCP}}
\newcommand{\nupdf}{\texorpdfstring{$\nu$}{nu}}
\newcommand{\nuenumupdf}{\texorpdfstring{$\nu_e/\nu_\mu$ }{nuenumu}}

\def\arrvline{\hfil\kern\arraycolsep\vline\kern-\arraycolsep\hfilneg}

\newcommand{\dcp}{\delta_{\rm CP}}
\newcommand{\nue}{\nu_e}

\newcommand{\anue}{\bar{\nu}_e}

\newcommand{\dm}[1]{\Delta m^2_{#1}}
\newcommand{\sinsq}[1]{\sin^2 \theta_{#1}}
\newcommand{\cossq}[1]{\cos^2 \theta_{#1}}

\newcolumntype{Y}{>{\centering\arraybackslash}X}

\definecolor{mycoral}{HTML}{FF7F50}

\allowdisplaybreaks

\let\OLDthebibliography\thebibliography
\renewcommand\thebibliography[1]{
\OLDthebibliography{#1}
\setlength{\parskip}{0pt}
\setlength{\itemsep}{0pt plus 0.3ex}
}

\begin{document}

\vspace{1cm}

\begin{titlepage}

\vspace{0.2truecm}

\begin{center}
\renewcommand{\baselinestretch}{1.8}\normalsize
\boldmath
{\LARGE\textbf{
Improving the robustness of the $\delta_{CP}$ determination with $\nu$SCOPE}}
\unboldmath
\end{center}

\vspace{0.4truecm}

\renewcommand*{\thefootnote}{\fnsymbol{footnote}}

\begin{center}

{

Jo\~ao Paulo Pinheiro$^{1,2}$\footnote{\href{mailto:joaopaulo.pinheiro@sjtu.edu.cn}{joaopaulo.pinheiro@sjtu.edu.cn}}\orcidlink{0000-0002-8777-9338}
and Salvador Urrea$^{3}$\footnote{\href{mailto:salvador.urrea@ijclab.in2p3.fr}{salvador.urrea@ijclab.in2p3.fr}}\orcidlink{0000-0002-7670-232X}
}

\vspace{0.7truecm}

{\footnotesize
$^1$State Key Laboratory of Dark Matter Physics, Tsung-Dao Lee Institute \& School of Physics and Astronomy, Shanghai Jiao Tong University, Shanghai 200240, China \\
$^2$Key Laboratory for Particle Astrophysics and Cosmology (MOE) \& Shanghai Key Laboratory for Particle Physics and Cosmology, Shanghai Jiao Tong University, Shanghai 200240, China
\\
$^3$ IJCLab, Pôle Théorie (Bat. 210), CNRS/IN2P3, 91405 Orsay, France
}

\vspace*{2mm}
\end{center}

\renewcommand*{\thefootnote}{\arabic{footnote}}
\setcounter{footnote}{0}

\begin{abstract}
The determination of leptonic CP violation is a primary goal of future long-baseline neutrino experiments such as DUNE and T2HK. The extraction of $\delta_{\mathrm{CP}}$ relies on the near-to-far extrapolation and on the assumed knowledge of the cross-section ratios $\sigma_{\nu_e}/\sigma_{\nu_\mu}$ and $\sigma_{\bar{\nu}_e}/\sigma_{\bar{\nu}_\mu}$, which are typically inferred under theoretical assumptions such as lepton universality and depend on nuclear modeling. In this work, we quantify how much of the sensitivity of DUNE and T2HK arises from these assumptions by performing a model-agnostic, data-driven estimation of systematic uncertainties in $\nu_e$ and $\bar{\nu}_e$ cross sections. We find that adopting such an agnostic approach can substantially degrade the CP-violation sensitivity, reducing it by nearly $3\sigma$ at maximal CP violation for DUNE, and $4\sigma$ for T2HK. We then assess the impact of the proposed $\nu$SCOPE experiment, which, through a combination of neutrino tagging and the Narrow-Band Off-Axis technique, can provide percent-level measurements of $\sigma_{\nu_\mu}$ and $\sigma_{\bar{\nu}_\mu}$ and constrain the ratio $\sigma_{\nu_e}/\sigma_{\nu_\mu}$ at the $\sim 2\%$ level. We show that including prospective $\nu$SCOPE measurements largely restores the lost sensitivity, highlighting that precise external cross-section measurements may be essential for a fully robust determination of $\delta_{\mathrm{CP}}$ and for breaking its degeneracy with nuclear mis-modeling or possible new physics affecting neutrino detection.
\end{abstract}

\end{titlepage}

\tableofcontents

\section{Introduction}

The discovery of neutrino oscillations confirmed that neutrinos have non-zero
masses and mix, providing the first robust evidence of physics beyond the Standard Model (BSM)~\cite{SNO:2002tuh,Super-Kamiokande:1998kpq,KamLAND:2002uet}. 
The mixing is parameterized by the leptonic mixing matrix, which contains three mixing angles 
($\theta_{12}, \theta_{13}, \theta_{23}$) and one Dirac CP-violating phase ($\dcp$).
While the mixing angles and mass splittings have been measured with increasing precision, the value of $\dcp$ remains one of the most challenging parameters to determine experimentally. A CP-violating value of $\dcp$ ($\dcp \neq 0, \pi$) would provide the first evidence of CP violation in the lepton sector and would establish the presence of an ingredient relevant to leptogenesis scenarios~\cite{Sakharov:1967dj} that aim to explain the observed baryon asymmetry of the Universe. It would not, however, demonstrate that leptogenesis generated this asymmetry. In standard leptogenesis scenarios, the relevant CP violation arises from complex phases associated with the couplings and decays of the heavy neutrino states, which are in general independent of the Dirac CP phase $\dcp$ of the low-energy PMNS matrix. Therefore, no model-independent connection can be established between the value of $\dcp$ measured in neutrino oscillations and the baryon asymmetry~\cite{Davidson:2008bu,Buchmuller:2004nz}.

Neutrino oscillation physics is entering a precision era in which the sensitivity of current and next-generation experiments is no longer limited by statistics, but by systematic uncertainties. In future long-baseline facilities—like the Deep Underground Neutrino Experiment (DUNE)~\cite{DUNE:2020ypp,DUNE:2020jqi} in the United States and the Tokai-to-Hyper-Kamiokande (T2HK) beam program, using the Hyper-Kamiokande (HK) detector, in Japan—the dominant contributions arise from hadron production, which governs neutrino flux predictions, and from neutrino--nucleus interaction modeling in the few-GeV regime~\cite{NuSTEC:2017hzk}. To mitigate these uncertainties, long-baseline experiments rely on near detector (ND) complexes, which, being located close to the source, provide high-statistics measurements of the unoscillated neutrino flux, with the aim of reducing the impact of systematic uncertainties on the far-detector (FD) analysis. The control of systematic uncertainties is of paramount importance for new physics searches~\cite{Coloma:2021uhq}, and is also essential for a robust determination of the Dirac CP-violating phase $\dcp$, whose measurement constitutes one of the flagship goals of these experiments, with a projected sensitivity reaching up to $5\sigma$ over a wide region of parameter space. 

Current-generation experiments already provide a concrete illustration of how systematic uncertainties may impact the determination of $\delta_{\text{CP}}$, with the notable tension between the results of T2K~\cite{T2K:2019bcf,T2K:2013ppw,T2K:2011ypd,T2K:2011qtm} and NOvA~\cite{NOvA:2004blv,NOvA:2021nfi}: T2K data favors maximal CP violation ($\dcp \approx -\pi/2$) and normal mass ordering (NO), while NOvA prefers values closer to CP conservation and exhibits a weaker preference for the mass ordering. Although this discrepancy may be a statistical fluctuation at the $\sim 2\sigma$ level~\cite{T2K:2025wet}, it could also point to unresolved systematic effects, including uncertainties in flux predictions  and neutrino--nucleus interaction modeling~\cite{Coyle:2025xjk}.

The measurement of $\dcp$ in the appearance channel $P_{\nu_{\mu}\to\nu_{e}}$ at the FD relies on ND data to constrain systematic uncertainties, and depends on the validity and accuracy of the ND-to-FD extrapolation. Oscillation experiments do not measure
probabilities directly, but event rates, in which fluxes, cross
sections, and detector effects enter in a convoluted manner and
cannot be disentangled with high precision, largely due to the
sizable uncertainties affecting both neutrino fluxes and
interaction cross sections. In the few-GeV regime relevant for
DUNE and T2HK, neutrino interactions arise from a combination of quasi-elastic (QE) scattering, resonance production (RES), and deep
inelastic scattering (DIS), all affected by nuclear final-state
interactions (FSI), whose mis-modeling—such as missing energy
carried by neutrons or sub-threshold hadrons—can bias the
reconstructed energy scale. At the ND, measurements probe the
unoscillated rate proportional to
$\phi_{\nu_{\mu}}\sigma_{\nu_{\mu}}$, whereas at the FD the $\nu_e$
appearance signal depends on $\phi_{\nu_{\mu}} P_{\nu_{\mu} \to
\nu_{e}} \sigma_{\nu_e}$; consequently, the ND-to-FD extrapolation
necessarily relies on theoretical assumptions about the ratio of
$\nu_e$ to $\nu_\mu$ cross sections,
$\sigma_{\nu_e}/\sigma_{\nu_\mu}$, invoking lepton universality and
nuclear modeling, to reduce the uncertainty at the level of $\sim
3\%$~\cite{Dieminger:2023oin}.

The reduction of systematic
uncertainties has motivated the development of strategies such as
the proposed PRISM
(Precision Reaction-Independent Spectrum Measurement)
strategy~\cite{Hasnip:2023ygr,Gehrlein:2025lbj,DUNE:2025lvs} for DUNE and the IWCD (Intermediate Water Cherenkov Detector)~\cite{nuPRISM:2014mzw} for T2HK, which measures the unoscillated flux at different off-axis angles. Beyond the mitigation strategies that can be implemented by each individual experiment, it is clear that a robust measurement of neutrino cross sections in the few-GeV region would be of enormous benefit for the future neutrino oscillation program. In this context, the proposed $\nu$SCOPE experiment at CERN~\cite{Acerbi:2025wzo} is particularly promising. Its design enables event-by-event tagging of the parent meson responsible for neutrino production, thereby providing unprecedented control over the neutrino flux. This would allow a measurement of $\sigma_{\nu_\mu}$ with an unprecedented bin-to-bin precision at the level of $2\%$. In addition, by using the Narrow-Band Off-Axis (NBOA) technique, whereby interactions are measured at different off-axis angles within the detector, $\nu$SCOPE is expected to determine the ratio $\sigma_{\nu_e}/\sigma_{\nu_\mu}$ with a precision of about $2\%$.

In this work, we aim to quantify how much of the sensitivity of DUNE and T2HK to $\delta_{\text{CP}}$ arises from the theoretical assumptions of lepton universality and nuclear modeling, and how much is independent of these assumptions. To do this, we perform a model-agnostic, data-driven estimation of systematic uncertainties in $\nu_e$ and $\bar{\nu}_e$ cross sections by using current data. Then, we assess the impact of these uncertainties on the determination of $\delta_{\text{CP}}$ in DUNE and T2HK. Finally, we study the prospective combination with $\nu$SCOPE in order to quantify its ability to deliver a robust determination of $\dcp$ and to disentangle genuine CP-violating effects from new physics in neutrino interactions, possible violations of lepton universality by new physics, and nuclear mis-modeling.

The remainder of this paper is organized as follows. In Sec.~\ref{sec:deltaCP_theory}, we review the challenges associated with a precision determination of $\dcp$. In Sec.~\ref{sec:cross_section_status}, we summarize the present experimental status of charged-current neutrino cross-section measurements for both the $\nu_\mu$ and $\nu_e$ sectors. In Sec.~\ref{sec:nuscope}, we present the main features of the proposed $\nu$SCOPE experiment. In Sec.~\ref{sec:DUNE_exp}, we describe the DUNE and T2HK experimental setup and their dominant systematic uncertainties. In Sec.~\ref{sec:analysis}, we detail our analysis, including the agnostic fit to current cross-section data, the implementation of data-driven systematics in DUNE and T2HK future sensitivities, and the inclusion of prospective $\nu$SCOPE measurements. Our main results are then presented in Sec.~\ref{sec:results}. Finally, we summarize our findings and conclude in Sec.~\ref{sec:summary_conclusions}. We also include two appendices giving additional results regarding the determination of the $\theta_{23}$ octant and the technical details of our implementation in the software GLoBES. 

\section{The Challenge of Precision \dcppdf Measurement}
\label{sec:deltaCP_theory}

In this section, we review the effects of CP violation in neutrino oscillations and the challenges in its determination. We begin by presenting the role of $\dcp$ in the oscillation probability in vacuum (Sec.~\ref{sec:dcp_oscillations}), then address the modifications induced by matter effects (Sec.~\ref{sec:matter_effects}), and finally discuss the experimental strategy based on near-to-far extrapolation and its sources of systematic uncertainty. We conclude with a discussion of the T2K--NO$\nu$A current tension in the $\dcp$ determination (Sec.~\ref{sec:T2K_Nova}).

\subsection{The leptonic CP-violating phase and its role in oscillations}
\label{sec:dcp_oscillations}

In the standard three-neutrino framework, the leptonic mixing matrix
contains three mixing angles ($\theta_{12}$, $\theta_{13}$,
$\theta_{23}$) and one irremovable Dirac CP-violating phase~$\dcp$.%
\footnote{If neutrinos are Majorana particles, two additional phases
appear; however, they are unobservable in oscillation experiments and
will not concern us here.}
CP violation in neutrino oscillations manifests as a difference between
neutrino and antineutrino transition probabilities,
$P(\nu_\alpha \to \nu_\beta) \neq P(\bar\nu_\alpha \to
\bar\nu_\beta)$ for $\alpha \neq \beta$, whose magnitude is controlled
by the Jarlskog invariant~\cite{Jarlskog:1985cw,Jarlskog:1985ht},
\begin{equation}
J = \frac{1}{8}\,
\sin 2\theta_{12}\,\sin 2\theta_{13}\,\sin 2\theta_{23}\,
\cos\theta_{13}\,\sin\dcp\,.
\label{eq:jarlskog}
\end{equation}
Since all three mixing angles are known to be non-zero, establishing
leptonic CP violation reduces to determining whether
$\dcp \neq 0, \pi$.

The sensitivity to $\dcp$ is concentrated in appearance channels since disappearance probabilities depend only on
$|U_{\alpha i}|^2$ and are therefore manifestly phase-independent.
In the $\nu_\mu \to \nu_e$ channel, expanding in
$\alpha \equiv \dm{21}/\dm{31} \approx 0.03$ and
$\sin\theta_{13} \approx 0.15$, the vacuum probability
reads~\cite{Akhmedov:2004ny}
\begin{align}
P(\nu_\mu \to \nu_e)
&\approx
\underbrace{4\,\cossq{13}\,\sinsq{13}\,\sinsq{23}\;\sin^2\!\Delta}
_{\text{leading CP-even}}
\nonumber\\[4pt]
&\quad
+\, 2\alpha\,\sin 2\theta_{13}\,\sin 2\theta_{12}\,\sin 2\theta_{23}\;
\sin\Delta
\bigl[\cos\Delta\,\cos\dcp - \sin\Delta\,\sin\dcp\bigr]
+ \mathcal{O}(\alpha^2)\,,
\label{eq:Pvac}
\end{align}
where $\Delta \equiv \dm{31}L/(4E)$.  Two features are worth highlighting. First, the CP-odd term (proportional to $\sin\dcp$) arises only through the interference between the atmospheric and solar oscillation amplitudes, and is therefore suppressed by a factor $\alpha$ relative to the leading contribution. Second, this suppression implies that extracting $\dcp$ from the event spectrum requires exquisite control over all multiplicative factors entering the observed rate, in particular the neutrino cross section (see Sec.~\ref{subsec:fd_nd}).

\subsection{Matter effects and their interplay with \dcppdf}
\label{sec:matter_effects}

As neutrinos traverse the Earth's mantle, electron neutrinos experience
the charged-current MSW
potential~\cite{Wolfenstein:1977ue,Mikheyev:1985zog}
$V_{\text{CC}} = \sqrt{2}\,G_F\,N_e$, whose effect on the appearance
probability is parameterized by
\begin{equation}
A = \frac{2\sqrt{2}\,G_F\,N_e\,E}{\dm{31}}\,.
\end{equation}
Since $A \to -A$ for antineutrinos, the matter medium itself breaks the
$\nu$/$\bar\nu$ symmetry, inducing a fake CP asymmetry even for
$\dcp = 0$.  Including constant-density matter effects, the appearance
probability becomes~\cite{Nunokawa:2007qh}
\begin{align}
P(\nu_\mu\to\nu_e)
&\approx
4\,\cossq{13}\,\sinsq{13}\,\sinsq{23}\;
\frac{\sin^2[(1-A)\Delta]}{(1-A)^2}
\nonumber\\[4pt]
&\quad
+\, 2\alpha\,\sin 2\theta_{13}\,\sin 2\theta_{12}\,\sin 2\theta_{23}\;
\frac{\sin(A\Delta)}{A}\,
\frac{\sin[(1-A)\Delta]}{1-A}
\nonumber\\
&\qquad\times
\bigl[\cos\Delta\,\cos\dcp - \sin\Delta\,\sin\dcp\bigr]
+ \mathcal{O}(\alpha^2)\,.
\label{eq:Pmat}
\end{align}
The structure mirrors the vacuum case, but now with $A$-dependent
modulation factors that differ between neutrinos and antineutrinos.
Because $A \propto E$ while $\Delta \propto 1/E$, the matter-induced
and genuine CP-violating asymmetries exhibit distinct energy
dependences, allowing their separation through spectral analysis ---
provided the beam covers a broad energy range.  This is a central
design feature of DUNE, whose wide-band beam spans roughly
$0.5$--$8\,\mathrm{GeV}$ over a $1300\,\mathrm{km}$ baseline, where
$A = \mathcal{O}(0.1\text{--}1)$.  At this baseline, matter effects
are also large enough to determine the mass ordering to a good
approximation, substantially reducing its degeneracy with~$\dcp$.

This spectral disentanglement, however, relies on the assumption that
the energy dependence of the observed event rate faithfully reflects
the oscillation probability.  As we discuss in
Sec.~\ref{subsec:fd_nd}, this assumption breaks down if the neutrino
cross section carries uncontrolled energy-dependent distortions.

\subsection{From event rates to transition probabilities: the role of
the near detector and lepton universality}
\label{subsec:fd_nd}

Long-baseline neutrino experiments employ high-intensity proton beams that, upon impinging on a target, produce secondary pions and kaons. These mesons are sign-selected by magnetic horns and decay in flight ($\pi^+ \to \mu^+ \nu_\mu$, $K^+ \to \mu^+ \nu_\mu$, and charge-conjugate modes), generating a nearly pure $\nu_\mu$ (or $\bar{\nu}_\mu$) beam with an energy spectrum peaked around the first oscillation maximum, $E \sim \dm{31} L/(2\pi)$. The neutrino flux is first characterized at a near detector (ND), located hundreds of meters from the source, and subsequently measured at a far detector (FD) positioned at baselines of hundreds to thousands of kilometers. The FD is designed to probe neutrino oscillations, in particular the appearance channel governed by $P(\nu_\mu \to \nu_e)$.  Theoretically, the CP-violating asymmetry
\begin{equation}
A_{\text{CP}}
\equiv
\frac{P(\nu_\mu\to\nu_e) - P(\bar\nu_\mu\to\bar\nu_e)}
{P(\nu_\mu\to\nu_e) + P(\bar\nu_\mu\to\bar\nu_e)}
\end{equation}
is directly sensitive to $\sin\dcp$ and vanishes when CP is
conserved.  In practice, however, oscillation experiments do not
measure $P(\nu_\mu \to \nu_e)$ directly --- they measure event rates.
The translation from rates to probabilities introduces a critical
dependence on neutrino cross sections, which we now discuss in detail.

The expected $\nu_e$ appearance event rate at the far detector as a function of reconstructed neutrino energy can be written as\footnote{To simplify the notation, we discuss neutrinos; the same discussion applies to antineutrinos. The overall normalization (number of targets, exposure) is omitted.}

\begin{equation}
\frac{dN^{\mathrm{FD}}_{\nu_e}}{dE^{\mathrm{rec}}}
\sim
\int dE^{\mathrm{true}}\;
\Phi^{\mathrm{FD}}_{\nu_\mu}(E^{\mathrm{true}})\,
P_{\mu e}(E^{\mathrm{true}})\,
\sigma_{\nu_e}(E^{\mathrm{true}})\,
R_{\nu_e}(E^{\mathrm{rec}},E^{\mathrm{true}})\,,
\label{eq:NFD_spectrum}
\end{equation}
where $\Phi^{\mathrm{FD}}_{\nu_\mu}$ is the unoscillated flux at the far detector, $\sigma_{\nu_e}$ the inclusive CC cross section, and $R_{\nu_e}$ the detector response function encoding efficiencies and energy smearing.

A near detector, placed sufficiently close to the source such that standard oscillations are negligible, measures the unoscillated event rate:
\begin{equation}
\frac{dN^{\mathrm{ND}}_{\nu_\alpha}}{dE^{\mathrm{rec}}}
\sim
\int dE^{\mathrm{true}}\;
\Phi^{\mathrm{ND}}_{\nu_\alpha}(E^{\mathrm{true}})\,
\sigma_{\nu_\alpha}(E^{\mathrm{true}})\,
R_{\nu_\alpha}(E^{\mathrm{rec}},E^{\mathrm{true}})\,,
\qquad \alpha = e,\,\mu\,.
\label{eq:NND}
\end{equation}
Since both the fluxes and cross sections are affected by large theoretical uncertainties, the near detector can only constrain their convolution, but not each factor independently. The oscillation probability is then inferred from the near-to-far event ratio,
\begin{equation}
\frac{dN^{\mathrm{FD}}_{\nu_e}/dE^{\mathrm{rec}}}
{dN^{\mathrm{ND}}_{\nu_\mu}/dE^{\mathrm{rec}}}
\;\sim\;
\frac{\Phi^{\mathrm{FD}}_{\nu_\mu}\;
P(\nu_\mu\to\nu_e)\;
\sigma_{\nu_e}}
{\Phi^{\mathrm{ND}}_{\nu_\mu}\;
\sigma_{\nu_\mu}}\,,
\label{eq:FDND_ratio}
\end{equation}
where a large fraction of correlated systematic uncertainties cancels. However, two sources of uncertainty remain irreducible at this stage.

\paragraph{Flux extrapolation.}  The ratio
$\Phi^{\mathrm{FD}}_{\nu_\mu}/\Phi^{\mathrm{ND}}_{\nu_\mu}$ is not
unity: the near detector subtends a much larger solid angle and
therefore samples a broader range of parent meson kinematics, leading
to a different neutrino energy spectrum.  A reliable extrapolation
requires good control of the flux at the near detector, which itself
depends on the cross-section modeling through Eq.~\eqref{eq:NND}.

\paragraph{The cross-section ratio
$\sigma_{\nu_e}/\sigma_{\nu_\mu}$.} Within the Standard Model, lepton universality
ensures identical weak couplings to the $W$ boson; differences between
$\sigma_{\nu_e}$ and $\sigma_{\nu_\mu}$ arise only from
charged-lepton mass effects --- phase-space suppression, contributions
from the pseudoscalar form factor ($\propto m_\ell^2$), and radiative
corrections --- typically at the few-percent
level~\cite{Nikolakopoulos:2019qcr,Dieminger:2023oin}.  In the
few-GeV regime relevant for DUNE, however, quasi-elastic, resonance,
and deep-inelastic processes coexist on nuclear targets, and
final-state interactions (FSI) redistribute visible energy through
neutrons, nuclear de-excitation, or absorbed pions.  Since FSI effects
depend on the outgoing charged-lepton kinematics, they induce an
additional flavor dependence beyond the purely electroweak level.
Under the combined assumptions of lepton universality and accurate
nuclear modeling, theoretical estimates place the uncertainty on
$\sigma_{\nu_e}/\sigma_{\nu_\mu}$ at about
$3\%$~\cite{Dieminger:2023oin}. 

In this work, we consider possible departures from this expectation being agnostic about their origin. Within the SM, such departures may arise from imperfect modeling of neutrino--nucleus interactions, including nuclear effects and final-state interactions, which can induce flavor- and energy-dependent distortions of the predicted cross sections.~\cite{Nikolakopoulos:2019qcr,Dieminger:2023oin} Beyond the SM, new interactions affecting neutrino detection could instead modify the electron- and muon-flavor cross sections differently~\cite{Kopp:2024yvh}, effectively violating the lepton-universality assumption used in the near-to-far extrapolation. Our analysis is designed to capture both possibilities through generic energy-dependent deformations of the electron-flavor cross sections.

\begin{figure}[t!]
\centering
\includegraphics[width=1\linewidth]{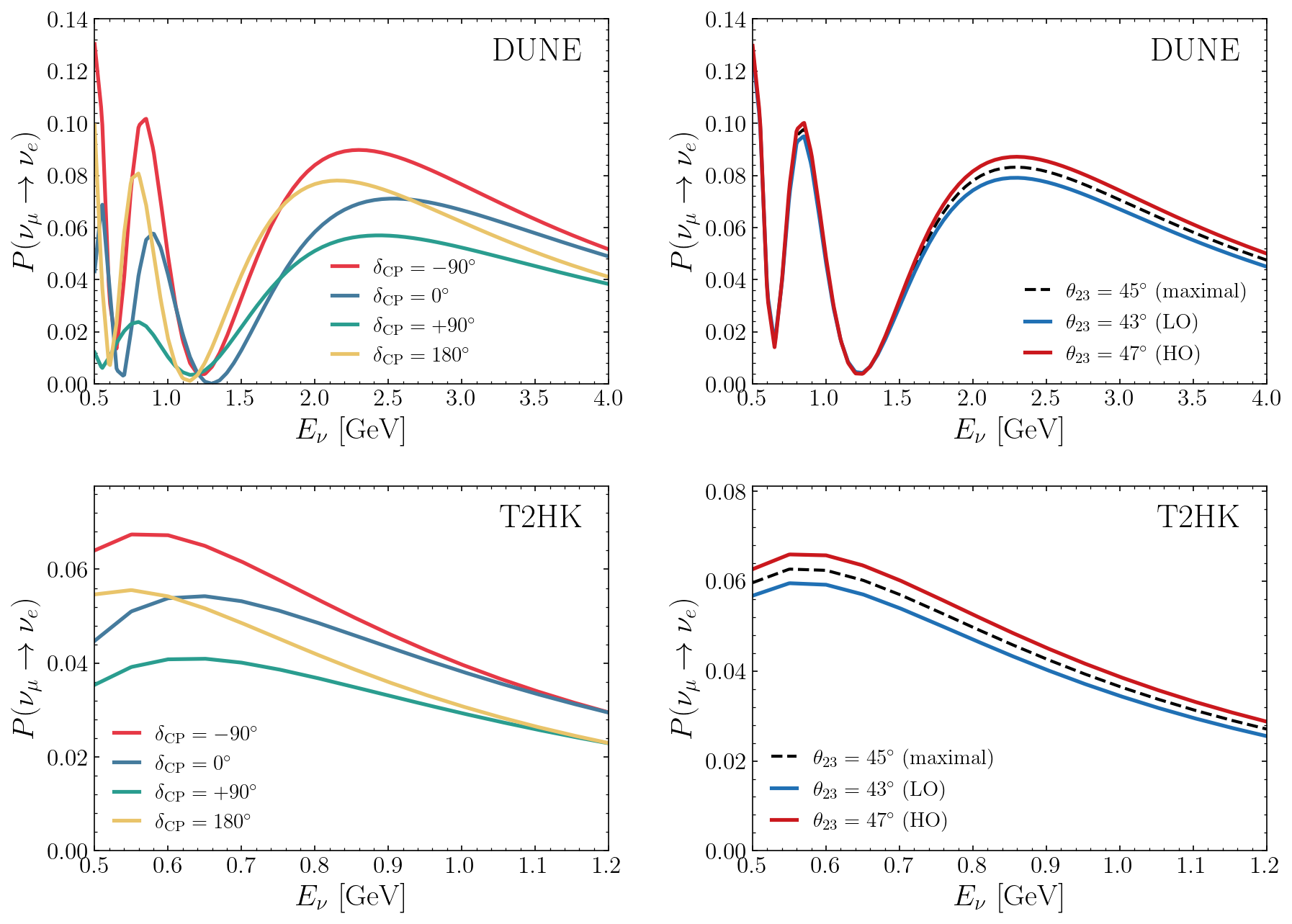}
\caption{Appearance probability $P(\nu_\mu \to \nu_e)$ as a 
function of neutrino energy at the DUNE baseline           
($L = 1300\,\mathrm{km}$, top row) and at the T2HK baseline 
($L = 295\,\mathrm{km}$, bottom row). \emph{Left panels:} $\theta_{23} = 48^\circ$ is fixed in the                  
higher octant and $\dcp$ varies over its full range; coloured                 
curves highlight $\dcp = -90^\circ$ (red), $0^\circ$ (blue),                  
$+90^\circ$ (green), and $180^\circ$ (yellow).
\emph{Right panels:} $\dcp = -90^\circ$ is fixed and $\theta_{23}$            
varies between $40^\circ$ and $50^\circ$; the dashed     
black line marks the maximal-mixing case
$\theta_{23} = 45^\circ$, while the blue and red curves highlight             
$\theta_{23} = 43^\circ$ (lower octant) and               
$\theta_{23} = 47^\circ$ (higher octant), respectively.     
All other oscillation parameters are fixed to the NuFIT~6.1 
best-fit values in normal ordering~\cite{Esteban:2024eli}.}
\label{fig:prob_panels}
\end{figure}

The danger for the $\dcp$ measurement of these sources of uncertainty can be stated as follows: as
shown in Eq.~\eqref{eq:Pvac}, the CP-violating modulation of
the appearance probability is a subleading, $\alpha$-suppressed
energy-dependent correction.  If these sources of uncertainty induce an uncontrolled, energy-dependent distortion whose spectral shape mimics that of $P(\nu_\mu \to \nu_e; \dcp)$ for some value of $\dcp$, such a distortion can become partially degenerate with the genuine CP-violating signal, thereby degrading or biasing the extracted value of $\dcp$.  In Figure~\ref{fig:prob_panels} we
show the appearance probability $P(\nu_\mu \to \nu_e)$ as a 
function of neutrino energy at the DUNE baseline (top row, 
$L = 1300\,\mathrm{km}$) and at the T2HK baseline (bottom row, 
$L = 295\,\mathrm{km}$). The left panels, obtained with 
$\theta_{23} = 48^\circ$ fixed in the higher octant, show the 
characteristic spectral modulation driven by $\dcp$; the right 
panels, obtained with $\dcp = -90^\circ$ fixed, show the 
complementary modulation driven by $\theta_{23}$ as it varies 
across the two octants. These are precisely the spectral 
features that any cross-section distortion would have to mimic 
in order to bias the oscillation analysis.

\subsection{The T2K--NO\nupdf A tension: a warning sign}
\label{sec:T2K_Nova}

The sources of uncertainty described above may already be affecting current neutrino data. T2K~\cite{T2K:2019bcf,T2K:2013ppw,T2K:2011ypd,T2K:2011qtm} and
NO$\nu$A~\cite{NOvA:2004blv,NOvA:2021nfi} have accumulated
substantial datasets in both neutrino and antineutrino modes, and a
mild but persistent $\sim 2\sigma$ tension
exists~\cite{T2K:2025wet} between their preferred regions in the
$\dcp$--$\theta_{23}$ plane: T2K favours maximal CP violation
($\dcp \approx -\pi/2$) with nearly maximal $\theta_{23}$ and normal
ordering, while NO$\nu$A prefers values closer to CP conservation.

This discrepancy may be a statistical fluctuation, but it may also
reflect uncontrolled systematics.  The two experiments differ in their
interaction generators (NEUT~\cite{Hayato:2021heg}
vs.\ GENIE~\cite{Andreopoulos:2015wxa,Tena-Vidal:2021rpu}), target
nuclei (oxygen vs.\ carbon), detector technologies (water Cherenkov
vs.\ liquid scintillator), and energy reconstruction methods
(kinematic vs.\ calorimetric), implying different sensitivities to
nuclear modeling and final-state interactions.  On the new-physics
side, non-standard interactions during
propagation~\cite{Chatterjee:2020kkm}
could introduce additional phases that become degenerate with $\dcp$
in a standard three-flavor fit or lead to tensions between different baseline experiments~\cite{Pinheiro:2026hmu}.

Regardless of its origin, this tension serves as a concrete reminder that statistical power alone may be insufficient. If driven by uncontrolled systematics, similar tensions could arise in the future. Experiments such as T2HK and DUNE are designed to achieve over $5\sigma$ sensitivity to CP violation across a wide range of parameter space; however, these projections rely on precise control of systematic uncertainties.  Quantifying this risk, and demonstrating how it can be mitigated through external cross-section measurements, is the central goal of the present work.

\section{Current status of charged-current cross-section measurements}
\label{sec:cross_section_status}

With the aim of estimating, in an agnostic and data-driven way, the level of systematic uncertainties in current neutrino charged-current cross sections, we review in this section the present experimental data in the energy range $E_\nu \in [0.5, 10]~\mathrm{GeV}$, which is most relevant for the next generation of long-baseline experiments. The landscape is strongly asymmetric between the muon and electron flavors.

\subsection{Muon Neutrino Sector}

For $\nu_\mu$ and $\bar\nu_\mu$, a substantial body of inclusive CC cross-section data exists, accumulated over decades by experiments 
operating at various beam energies and on different nuclear targets. Modern measurements from MINERvA~\cite{MINERvA:2019gsf,MINERvA:2017btl}, 
T2K (using both INGRID and ND280 detectors)~\cite{T2K:2016sji,T2K:2017mci,T2K:2018rhz}, NOvA~\cite{NOvA:2024zkm,NOvA:2021nfi}, and 
MicroBooNE~\cite{MicroBooNE:2019npw,MicroBooNE:2024jjp,MicroBooNE:2022tur} have provided increasingly precise determinations of 
$\sigma_{\nu_\mu}/E$ on carbon, hydrocarbon, water, and argon, typically reaching uncertainties of $5$--$10\%$ per energy bin in the few-GeV 
range. These data, combined with older measurements from MINOS~\cite{MINOS:2010}, NOMAD~\cite{NOMAD:2008cc}, SciBooNE~\cite{SciBooNE:2011},
ArgoNeuT~\cite{ArgoNeuT:2011bms,ArgoNeuT:2014yyp}, 
NINJA~\cite{NINJA:2021},  NuTeV~\cite{NuTeV:2006}, CCFR~\cite{CCFR:1997}, IHEP-JINR~\cite{IHEP-JINR:1996}, CDHS~\cite{CDHS:1987}, 
BNL~\cite{BNL:1982}, Gargamelle (SPS/PS)~\cite{GGM-SPS:1981,GGM-PS:1979}, ANL~\cite{ANL:1979}, BEBC~\cite{BEBC:1979}, IHEP-ITEP~\cite{IHEP-ITEP:1979}, and SKAT~\cite{SKAT:1979},
constrain the $\nu_\mu/\bar{\nu}_\mu$ cross section
reasonably well across the GeV range, although 
tensions between generators remain at the level of the nuclear modeling uncertainties.

\subsection{Electron Neutrino Sector}

The situation for $\nu_e$ and $\bar\nu_e$ is dramatically worse. Historically, the only experiments that have measured inclusive CC $\nu_e$ 
cross sections with meaningful statistics in the GeV range are Gargamelle~\cite{Gargamelle:1977pmg,GGM-PS:1979} and a handful of bubble-chamber 
experiments from the 1970s and 1980s, such as BEBC~\cite{BEBC:1979}, SKAT~\cite{SKAT:1979}, and the 12-ft/15-ft chambers at 
ANL/BNL~\cite{ANL:1979,BNL:1982}. Remarkably, the Gargamelle data, collected nearly fifty years ago using a heavy-liquid bubble chamber at CERN,
still provide the most statistically significant historical constraints on $\sigma_{\nu_e}/E$ in the $1$--$10\,\mathrm{GeV}$ range. The 
$\bar\nu_e$ situation is even more limited: the world dataset in this energy range amounts to only a few hundred events, again dominated by 
Gargamelle.
Modern experiments such as T2K~\cite{T2K:2014nfh} and MicroBooNE~\cite{MicroBooNE:2021ccs,MicroBooNE:2022wqi} measured a small number
of data points for $\nu_e/\bar{ \nu_e}$ for fixed values of energy.

The scarcity of $\nu_e$ data has a simple experimental origin. Conventional neutrino beams are designed to produce $\nu_\mu$ from two-body pion
and kaon decays; the $\nu_e$ component is an intrinsic contamination arising predominantly from three-body decays ($\mu^+ \to e^+ \nu_e 
\bar\nu_\mu$, $K^+ \to \pi^0 e^+ \nu_e$), which typically contributes at the percent level of the total beam flux. As already discussed in Sec.~\ref{subsec:fd_nd}, cross sections and fluxes are measured together in event rates, and disentangling them is a difficult task. Therefore, extracting a $\nu_e$ cross section from these events requires precise knowledge of the $\nu_e$ flux, which is itself subject to large systematic uncertainties arising from the modeling of hadron production. 

More recent experiments have begun to improve this picture. T2K has measured the $\nu_e$ CC inclusive cross section on carbon at 
ND280~\cite{T2K:2014nfh}, and MicroBooNE has provided the first $\nu_e$ measurements on argon~\cite{MicroBooNE:2021ccs,MicroBooNE:2022wqi}, 
which is the target nucleus of direct relevance for DUNE. However, these measurements are still statistically limited (of order $10^3$--$10^4$ 
events), often cover restricted phase spaces, and carry substantial flux uncertainties inherited from the parent beam simulation.

No experiment has yet measured $\sigma_{\nu_e}/E$  with the percent-level precision needed for future-generation oscillation analyses. This is the gap that the proposed $\nu$SCOPE experiment, described in Sec.~\ref{sec:nuscope}, is designed to fill.

\section{\nupdf SCOPE}\label{sec:nuscope}

The neutrino SPS Complex for Precision Experiments ($\nu$SCOPE)~\cite{Acerbi:2025wzo} is a proposed short-baseline neutrino facility at CERN, specifically designed to eliminate the dominant sources of systematic uncertainties in neutrino cross-section measurements. It aims to support the next generation of long-baseline oscillation experiments by measuring neutrino interactions at the GeV scale with high precision. The beamline uses the 400~GeV/c proton beam from the Super Proton Synchrotron (SPS) to strike a graphite target, producing a secondary beam of kaons and pions with a central momentum of roughly 8.5~GeV/c. These mesons are subsequently focused and decay within a 40~m-long decay tunnel to produce a highly controlled neutrino beam. A 500-ton fiducial-mass liquid argon time projection chamber (LArTPC) will be positioned 25~m downstream of the decay tunnel.

What makes $\nu$SCOPE fundamentally different from conventional neutrino facilities is its ability to disentangle with high accuracy the neutrino flux and cross section. Building on the achievements of the NP06/ENUBET~\cite{ENUBET:2023hgu,Longhin:2014yta} and NuTag collaborations~\cite{Baratto-Roldan:2024bxk,Perrin-Terrin:2021jtl}, the decay tunnel is instrumented with calorimeters and photon vetoes, complemented by an instrumented hadron dump. This setup allows for the direct measurement of charged leptons (positrons and muons) produced in association with neutrinos during meson decays. By monitoring these leptons at the single-particle level, $\nu$SCOPE can constrain the inclusive neutrino flux normalization to the $1\%$ level, circumventing the standard $\sim 10\%$ uncertainty associated with conventional beam simulations. Knowing the flux with high precision, together with the implementation of neutrino tagging, enables a very precise measurement of $\sigma_{\nu_{\mu}}$ and $\sigma_{\bar\nu_{\mu}}$ (at the $\sim 1\%$ level in the $1$--$4~\mathrm{GeV}$ region, in $200~\mathrm{MeV}$-wide bins), as well as a precise determination of $\sigma_{\nu_e}/\sigma_{\nu_\mu}$ at the $\sim 2\%$ level using the Narrow-Band Off-Axis (NBOA) technique. The details of both neutrino tagging and NBOA will be presented next.  

\subsection{Neutrino tagging}\label{sec:neutrino_tagging}

Beyond monitoring the inclusive flux, $\nu$SCOPE incorporates advanced silicon pixel spectrometers to perform neutrino ``tagging'', a technique pioneered by the NuTag collaboration. The beamline is instrumented with a beam spectrometer upstream of the decay tunnel to precisely measure the trajectory and momentum of the parent mesons ($K^+, \pi^+$), and a muon spectrometer downstream of the decay tunnel to track the daughter muons resulting from two-body decays ($\pi^+ \to \mu^+ \nu_\mu$ and $K^+ \to \mu^+ \nu_\mu$). 

By utilizing fast, radiation-hard 4D silicon tracking technologies, these spectrometers achieve time resolutions of less than 100~ps. When combined with a fast photon detection system in the LArTPC (providing $\sim 300$~ps resolution), individual neutrino interactions in the detector can be uniquely matched in space and time to their parent meson decay in the tunnel.

For successfully tagged interactions, the true energy of the incoming neutrino can be kinematically reconstructed strictly from the two-body decay kinematics: $E_\nu = E_{\pi,K} - E_\mu$. This technique provides an \textit{a priori} measurement of the neutrino energy with a sub-percent resolution, entirely independent of the complex final-state particle reconstruction inside the nuclear medium of the LArTPC. Consequently, tagging allows for direct and unbiased measurements of neutrino cross-sections as a function of true neutrino energy and provides a unique method to precisely calibrate the energy reconstruction bias that limits current long-baseline analyses.

With the neutrino-tagging technique, the total statistics at each neutrino energy allow the extraction of $\sigma_{\nu_{\mu}},\sigma_{\bar{\nu}_{\mu}} $ with statistical uncertainties below $5\%$ across the entire considered energy range of $0$--$8$~GeV. The statistical uncertainty is expected to be below $\sim 1\%$ in the $1$--$4$~GeV region, where most neutrinos originate from pion decays and the flux peaks around $3$~GeV. A second peak appears around $7$~GeV, dominated by neutrinos from kaon decays, where a $2$--$3\%$ uncertainty in the cross-section determination is expected~\cite{Acerbi:2025wzo}. The worst sensitivity is found in the intermediate-energy region between the pion- and kaon-dominated fluxes, although even there the uncertainty is expected to remain below $5\%$.

While neutrino tagging provides a powerful handle on $\nu_\mu,\bar{\nu}_\mu$ from two-body decays, its extension to the $\nu_e$ sector is intrinsically more challenging. Electron neutrinos in the beam originate predominantly from three-body kaon decays ($K^+ \to \pi^0 e^+ \nu_e$), where the presence of an additional final-state particle prevents a full kinematic reconstruction of the neutrino energy on an event-by-event basis. As a consequence, the direct tagging technique cannot provide the same level of \textit{a priori} energy determination as in the two-body case. Instead, the electron neutrino sector will be constrained through the detection of associated positrons in the decay tunnel, and complemented by techniques such as Narrow-Band Off-Axis (NBOA), which we discuss next.

\subsection{Measurement of the \nuenumupdf 
cross-section ratio}

A precise determination of the $\nu_e/\nu_\mu$ cross-section ratio is required for an accurate extraction of $P_{\mu e}$, and ultimately to reach the desired sensitivity to $\delta_{\text{CP}}$ in future long-baseline experiments, as discussed in Sec.~\ref{subsec:fd_nd}. Unlike the muon sector, the $\nu_e$ and $\bar{\nu}_e$ components of the beam are produced predominantly through three-body decays (such as $K_{e3} \equiv K^+ \to \pi^0 e^+ \nu_e$), which makes event-by-event kinematic tagging challenging, as outlined in Sec.~\ref{sec:neutrino_tagging}.

To overcome this limitation, $\nu$SCOPE employs the Narrow-Band Off-Axis (NBOA) technique combined with the PRISM methodology~\cite{nuPRISM:2014mzw}. Due to the proximity and large transverse size of the LArTPC ($4\times4~\text{m}^2$), the detector spans an angular range up to $4.5^\circ$ off-axis. The $\nu_\mu$ flux, primarily originating from two-body decays, exhibits a strong correlation between its emission angle and energy. By dividing the detector into radial off-axis slices indexed by $j$, one effectively measures $\nu_\mu$ interactions exposed to a variety of distinct, narrow-band fluxes, which can be viewed as an approximate basis $\{\Phi^{\nu_{\mu}}_j(E_{\nu})\}$ of the functional space of fluxes.

Using linear combinations of these off-axis $\nu_\mu$ fluxes, one can construct a virtual flux that matches a desired ``target'' flux. We take the monitored $\nu_e$ flux measured by $\nu$SCOPE as this target:
\begin{equation}
\Phi^{\nu_e}(E_{\nu}) \simeq \sum_{j} c_{j} \Phi^{\nu_{\mu}}_j(E_{\nu}) \,,
\end{equation}
where the coefficients $c_{j}$ are obtained using Tikhonov~\cite{tikhonov1977solutions,Phillips:1962ofa}
regularization to control statistical fluctuations. This allows us to construct the corresponding virtual $\nu_\mu$ observable:
\begin{equation}
N^{\nu_{\mu}}_{\text{virtual}} = \int dE_{\nu} \, \Phi^{\nu_e}(E_{\nu}) \, \sigma_{\nu_{\mu}}(E_{\nu}) \;\simeq\; \sum_{j} c_{j} \, N^{\nu_{\mu}}_j \,,
\end{equation}
where $N^{\nu_{\mu}}_j$ denotes the number of $\nu_\mu$ events measured in the $j$-th radial off-axis slice.

The key advantage of this approach is that it enables an experimental determination of the cross-section ratio evaluated over the same flux shape:
\begin{equation}\label{eq:RNBOA}
R_{\text{NBOA}}
\equiv
\frac{N^{\nu_e}}{N^{\nu_\mu}_{\mathrm{virtual}}}
\;\simeq\;
\frac{
\displaystyle \int dE_\nu \, \Phi^{\nu_e}(E_\nu)\, \sigma_{\nu_e}(E_\nu)
}{
\displaystyle \int dE_\nu \, \Phi^{\nu_e}(E_\nu)\, \sigma_{\nu_\mu}(E_\nu)
}\,.
\end{equation}
This observable is significantly more constraining than a naive $N^{\nu_e}/N^{\nu_\mu}$ ratio, which depends on both $\Phi^{\nu_e}$ and $\Phi^{\nu_\mu}$ and therefore leaves additional freedom for cross-section distortions. By weighting both cross sections with the same $\Phi^{\nu_e}$ target flux, theoretical uncertainties associated with their energy dependence largely cancel. $\nu$SCOPE is expected to measure $R_{\text{NBOA}}$ with a statistical uncertainty of $\sim 2\%$ and a flux-related systematic uncertainty on $\Phi^{\nu_e}(E_\nu)$ at the $\sim 1\%$ level.

Finally, the same NBOA matching procedure could in principle be extended to antineutrinos by operating the SPS beamline in reversed polarity and focusing $K^-$ and $\pi^-$ mesons. Ref.~\cite{Acerbi:2025wzo} does not provide a dedicated sensitivity study for antineutrino running, although it states that similar measurements could be performed in this mode. In our analysis, we therefore assume, as a benchmark, the same $2\%$ precision as for neutrinos.
\section{Long Baseline Experiments}
\label{sec:DUNE_exp}

In this section we describe the experimental setups of DUNE and T2HK and identify
the systematic uncertainties most relevant for the $\dcp$
determination, highlighting those that a $\nu$SCOPE measurement would
directly address.

\subsection{Experimental setup}

\subsubsection{DUNE}

The DUNE neutrino beam, delivered by the Long-Baseline Neutrino
Facility (LBNF)~\cite{DUNE:2020fgq,DUNE:2020ypp,DUNE:2020jqi}, is
produced at Fermilab by directing $120\,\mathrm{GeV}$ protons onto a
segmented graphite target of length $1.5\,\mathrm{m}$.  The resulting
secondary hadrons are focused by a system of three magnetic horns and
enter a decay region of approximately $194\,\mathrm{m}$, where they
decay in flight to produce neutrinos.  Downstream of the decay pipe,
an absorber removes residual hadrons and unreacted protons, while a
muon shield suppresses charged contamination, yielding a predominantly
neutrino beam.

The horn polarity controls the beam composition: focusing $\pi^+/K^+$
(Forward Horn Current, FHC) produces a neutrino-enhanced beam, while
focusing $\pi^-/K^-$ (Reverse Horn Current, RHC) yields an
antineutrino-enhanced beam.  In FHC mode, the flux is dominated by
$\nu_\mu$ from $\pi^+$ and $K^+$ decays, peaking at
$2$--$3\,\mathrm{GeV}$, with a subdominant $\nu_e$ component arising
from muon decays near the peak and from $K^+$ three-body decays in the
high-energy tail.  A residual $\bar\nu_\mu$ contamination arises from
$\pi^-$ that are not fully defocused, while $\bar\nu_e$ are mainly
produced in $K_L^0$ decays.  In RHC mode the roles of neutrinos and
antineutrinos are reversed.

The Far Detector~(FD), located $1295\,\mathrm{km}$ away at the Sanford
Underground Research Facility, will consist of four liquid argon time
projection chamber (LArTPC) modules with a total fiducial mass of
$4\times 17\,\mathrm{kton}$.  The Near Detector~(ND) complex, located
$574\,\mathrm{m}$ from the target, includes a LArTPC (ND-LAr) with a
fiducial mass of $67.2\,\mathrm{tons}$ and the same nuclear target as
the FD, complemented by a muon spectrometer and the magnetized on-axis
detector SAND.  Phase~I is expected to begin operation around 2031 with
a beam power of $1.2~\mathrm{MW}$
($1.1\times10^{21}\,\mathrm{POT/yr}$), followed by a Phase~II upgrade
to $2.4\,\mathrm{MW}$, which may also include the addition of a
high-pressure gaseous argon TPC (ND-GAr) with enhanced tracking and
charge identification capabilities.

\subsubsection{T2HK}

The Tokai-to-Hyper-Kamiokande~(T2HK) experiment~\cite{Abe:2018uyc} employs a neutrino beam produced at the J-PARC facility in Tokai, Japan, and directed toward the Hyper-Kamiokande water-Cherenkov far detector, located $295,\mathrm{km}$ away in Kamioka.  The beam is generated by $30\,\mathrm{GeV}$ protons from the Main Ring synchrotron impinging on a graphite target, producing secondary pions and kaons that are focused by magnetic horns and subsequently decay in flight inside a dedicated decay volume.  The resulting flux is dominated by $\nu_\mu$ in neutrino mode (Forward Horn Current, FHC) and by $\bar\nu_\mu$ in antineutrino mode (Reverse Horn Current, RHC).  By placing the detector $2.5^\circ$ off the beam axis, the narrow off-axis beam technique yields a neutrino flux that peaks sharply at approximately $0.6\,\mathrm{GeV}$, close to the first oscillation maximum at this baseline, thereby maximizing the sensitivity to $\nu_\mu \to \nu_e$ transitions while suppressing the high-energy tail.

The Far Detector~(FD), housed in Tochibora mine in Kamioka, is a cylindrical water-Cherenkov tank with a fiducial mass of $187\,\mathrm{kton}$, approximately 8.4 times larger than its predecessor Super-Kamiokande~\cite{Abe:2015zbg}.  Neutrino interactions are detected through Cherenkov light emitted by final-state charged particles, and single-ring versus multi-ring topologies are used to discriminate $\nu_e$-like ($e$-ring) from $\nu_\mu$-like ($\mu$-ring) charged-current events.  The Near Detector~(ND) system includes the upgraded ND280 off-axis tracker at approximately $280\,\mathrm{m}$ from the target, which constrains the unoscillated beam flux and interaction rates, together with a new Intermediate Water Cherenkov Detector~(IWCD) at approximately $1\,\mathrm{km}$ from the source~\cite{hkNear}.  The IWCD shares the same water-Cherenkov technology and oxygen nuclear target as the FD, enabling more direct control of reconstruction efficiencies and detector-related systematic uncertainties compared to a ND with a different target medium.

\subsection{Systematic uncertainties and the role of \nupdf SCOPE}

The determination of $\dcp$ at DUNE and T2HK is expected to be limited primarily by
systematic rather than statistical uncertainties~\cite{Coloma:2021uhq}.
Among the most relevant are uncertainties in the neutrino flux, interaction
cross sections, and energy reconstruction. In both experiments, the far-detector
appearance signal depends on $\nu_e$ and $\bar\nu_e$ charged-current interactions,
whereas the near detector is dominated by $\nu_\mu$ and $\bar\nu_\mu$ events.
As discussed in Sec.~\ref{subsec:fd_nd}, this implies that the near-to-far
extrapolation depends critically on the cross-section ratios
$\sigma_{\nu_e}/\sigma_{\nu_\mu}$ and
$\sigma_{\bar\nu_e}/\sigma_{\bar\nu_\mu}$.
Under the assumptions of lepton universality and accurate nuclear modeling,
these ratios are usually taken to be known at the few-percent
level~\cite{Nikolakopoulos:2019qcr,Dieminger:2023oin}. However, as reviewed in
Sec.~\ref{sec:cross_section_status}, current direct data on the electron-flavor
cross sections remain too sparse to validate this assumption in a fully
data-driven way.

This limitation should be distinguished from other important sources of
systematic uncertainty that have motivated dedicated mitigation strategies within
the long-baseline program. For instance, Ref.~\cite{Coyle:2025xjk} showed that
mismodelling of final-state interactions can bias the
$E_\nu^{\mathrm{true}} \to E_\nu^{\mathrm{reco}}$ migration matrices, and that
on-axis near-detector tuning alone is insufficient to remove this effect,
thereby motivating DUNE-PRISM as a tool to constrain energy-reconstruction
systematics. Likewise, Ref.~\cite{Gehrlein:2025lbj} showed that imperfect
knowledge of the relative normalization of the different beam components can
bias the PRISM coefficients and reduce the CP-violation sensitivity, motivating
the Lateral Extraction of Neutrino Spectra (LENS) strategy to constrain the flux composition.

While these approaches are highly valuable, they do not directly address the
central vulnerability studied in this work: the poor experimental control of
$\sigma_{\nu_e}/\sigma_{\nu_\mu}$ and
$\sigma_{\bar\nu_e}/\sigma_{\bar\nu_\mu}$. Since PRISM- and LENS-like strategies
exploit the variation of the flux with off-axis angle, they constrain the beam
and reconstruction model, but not the flavor ratio of charged-current cross
sections itself. A measurement at $\nu$SCOPE would provide precisely this missing
input in two complementary ways. First, percent-level measurements of
$\sigma_{\nu_\mu}$ and $\sigma_{\bar\nu_\mu}$ in the few-GeV range would improve
the data-driven determination of the unoscillated flux and strengthen the
consistency of the
$\Phi^{\mathrm{FD}}_{\nu_\mu}/\Phi^{\mathrm{ND}}_{\nu_\mu}$ extrapolation.
Second, the measurement of $R_{\text{NBOA}}$ defined in
Eq.~\eqref{eq:RNBOA} would provide a direct constraint on
$\sigma_{\nu_e}/\sigma_{\nu_\mu}$ and
$\sigma_{\bar\nu_e}/\sigma_{\bar\nu_\mu}$.

For DUNE, this information is especially relevant because the few-GeV region
contains overlapping quasi-elastic, resonance, and deep-inelastic contributions,
while final-state interactions can significantly distort the visible-energy
spectrum in the liquid-argon detector. In this environment, external
cross-section information is particularly valuable to reduce the model dependence
of the appearance prediction.

T2HK faces the same conceptual issue, although in a different experimental
regime. Its water-Cherenkov far detector and narrow-band beam at
$0.6\,\mathrm{GeV}$ place the analysis closer to the quasi-elastic region, where
the interaction physics is simpler than in DUNE. Nevertheless, the extraction of
the appearance signal still depends on the same flavor ratios
$\sigma_{\nu_e}/\sigma_{\nu_\mu}$ and
$\sigma_{\bar\nu_e}/\sigma_{\bar\nu_\mu}$, which are not directly measured by the
$\nu_\mu$-dominated near detector. A precise determination of
$R_{\mathrm{NBOA}}$ at $\nu$SCOPE would therefore also be highly beneficial for
T2HK.

We note that $\nu$SCOPE is currently proposed to measure cross sections on argon, whereas T2HK employs a water target. However, the
dominant source of uncertainty constrained by $R_{\mathrm{NBOA}}$
--- namely, the electroweak $\nu_e/\nu_\mu$ cross-section ratio ---
is nucleus-independent at leading order: it is governed by the
charged-lepton mass difference and radiative corrections, which are
identical for all nuclear targets.  Nuclear corrections to this ratio,
arising from differences in Pauli blocking, binding energy, and
final-state interactions between $^{40}$Ar and $^{16}$O, are expected
to be subleading compared to the current level of uncertainty on the
ratio itself~\cite{Nikolakopoulos:2019qcr,Dieminger:2023oin}.  A
dedicated study of the nuclear-target dependence of these corrections
is beyond the scope of this work. 
\section{Analysis}\label{sec:analysis}

In this section, we detail our statistical analysis. Subsection~\ref{sec:cross_section} presents the fit to current neutrino cross-section data, while Subsection~\ref{sec:globes_implementation} discusses the implementation of data-driven systematic uncertainties, along with our treatment of a prospective $\nu$SCOPE measurement.

\subsection{Fit to current neutrino cross-section data}\label{sec:cross_section}

The purpose of this subsection is to obtain an agnostic, data-driven
estimate of the neutrino cross-section systematic uncertainties using
currently available data summarized in
Sec.~\ref{sec:cross_section_status}, covering the energy range
$E \in [0.4,\, 10.0]~\mathrm{GeV}$ relevant for DUNE and T2HK.

\subsubsection{Generation of cross-section deformations and \chipdf goodness of fit}\label{sec:distorsions}

As discussed in Sec.~\ref{subsec:fd_nd}, distortions in the energy dependence of the cross section $\sigma_{\nu_e}(E)$ that mimic the spectral shape of the appearance probability $P(\nu_\mu \to \nu_e;\,\dcp)$ can become partially degenerate with the genuine CP-violating signal. The aim of this subsection will be to quantify the class of cross-section deformations compatible with current data.

To this end, we construct a reference cross section by fitting the global dataset compiled in Sec.~\ref{sec:cross_section_status}. For each neutrino flavor $\alpha = \nu_e,\, \bar\nu_e,\, \nu_\mu,\, \bar\nu_\mu$, we define a smooth interpolation $(\sigma_\alpha / E)^{\mathrm{interp}}(E)$ that reproduces the experimental measurements. Deviations from this baseline are parametrized through an energy-dependent deformation $f_\alpha(E)$~ as
\begin{equation}
\left(\frac{\sigma_{\alpha}}{E}\right)^{\mathrm{def}}(E,f_{\alpha})
=
\left[1 + f_{\alpha}(E)\right]
\left(\frac{\sigma_{\alpha}}{E}\right)^{\mathrm{interp}}.
\end{equation}

Note that the cross-section deformations are, in principle, fully general. In practice, rather than scanning the full functional space, we use the strategy described in Appendix~\ref{App:deformation_details} to efficiently identify the deformations that are most relevant for the $\delta_{\rm CP}$ determination. Since some
experimental measurements are provided in energy bins with sizable
widths, the theoretical prediction is evaluated as a bin-averaged
quantity:
\begin{equation}
\left(\frac{\sigma_\alpha}{E}\right)^{\mathrm{avg}}_i (f_\alpha)
=
\frac{1}{\Delta E_i}
\int_{E_i^{\mathrm{min}}}^{E_i^{\mathrm{max}}}
dE \;
\left(\frac{\sigma_\alpha}{E}\right)(E, f_\alpha)\,,
\end{equation}
where $\Delta E_i = E_i^{\mathrm{max}} - E_i^{\mathrm{min}}$ denotes
the width of the $i$-th bin. The goodness-of-fit statistic for each
flavor channel is then defined as
\begin{equation}
\chi^2_{\alpha}(f_{\alpha}) =
\sum_{i=1}^{N}
\left(
\frac{
(\sigma_\alpha/E)_i^{\mathrm{data}} -
(\sigma_\alpha/E)_i^{\mathrm{avg}}(f_\alpha)
}{
\delta\sigma_i
}
\right)^2,
\label{eq:chisq_xsec}
\end{equation}
where $\delta\sigma_i$ denotes the experimental uncertainty in each
bin, and the sum runs over all $N$ data points. The statistic $\chi^2_{\alpha}$ approximately follows a $\chi^2$ distribution with $N$ degrees of freedom. Here, measurements from different experiments are treated as statistically independent. Many of the $\nu_e$ and $\bar{\nu}_e$ datasets consist of only a small number of energy bins, which reduces the impact of bin-to-bin correlations. For simplicity, and due to the lack of publicly available covariance matrices, correlations within each dataset are neglected.

In what follows, we focus on observables that are primarily determined by appearance
channels such as $\delta_{\text{CP}}$. Possible distortions in the disappearance channels are
expected to be much better constrained by near detectors as discussed in Sec.~\ref{subsec:fd_nd};
therefore, we assume that the dominant systematic uncertainty arises
from $\sigma_{\nu_e}$ and $\sigma_{\bar{\nu}_e}$ . For this reason, and to make our scans more
manageable, we restrict our analysis to distortions in the electron
sector. Also, the $\nu_e$ and $\bar\nu_e$ deformations are treated independently throughout the analysis. This is a conservative choice, since imposing a correlation between the two would introduce an additional theoretical assumption and could artificially strengthen the projected sensitivity. \footnote{We have checked, for a couple of values of $\delta_{CP}$, namely $\delta_{CP}=-100^\circ$ and $100^\circ$, whether distortions in $\sigma_{\mu}$ modify the results, and found that they do not have a noticeable impact on the results presented here.
}

\begin{figure}
\centering
\includegraphics[width=1\linewidth]{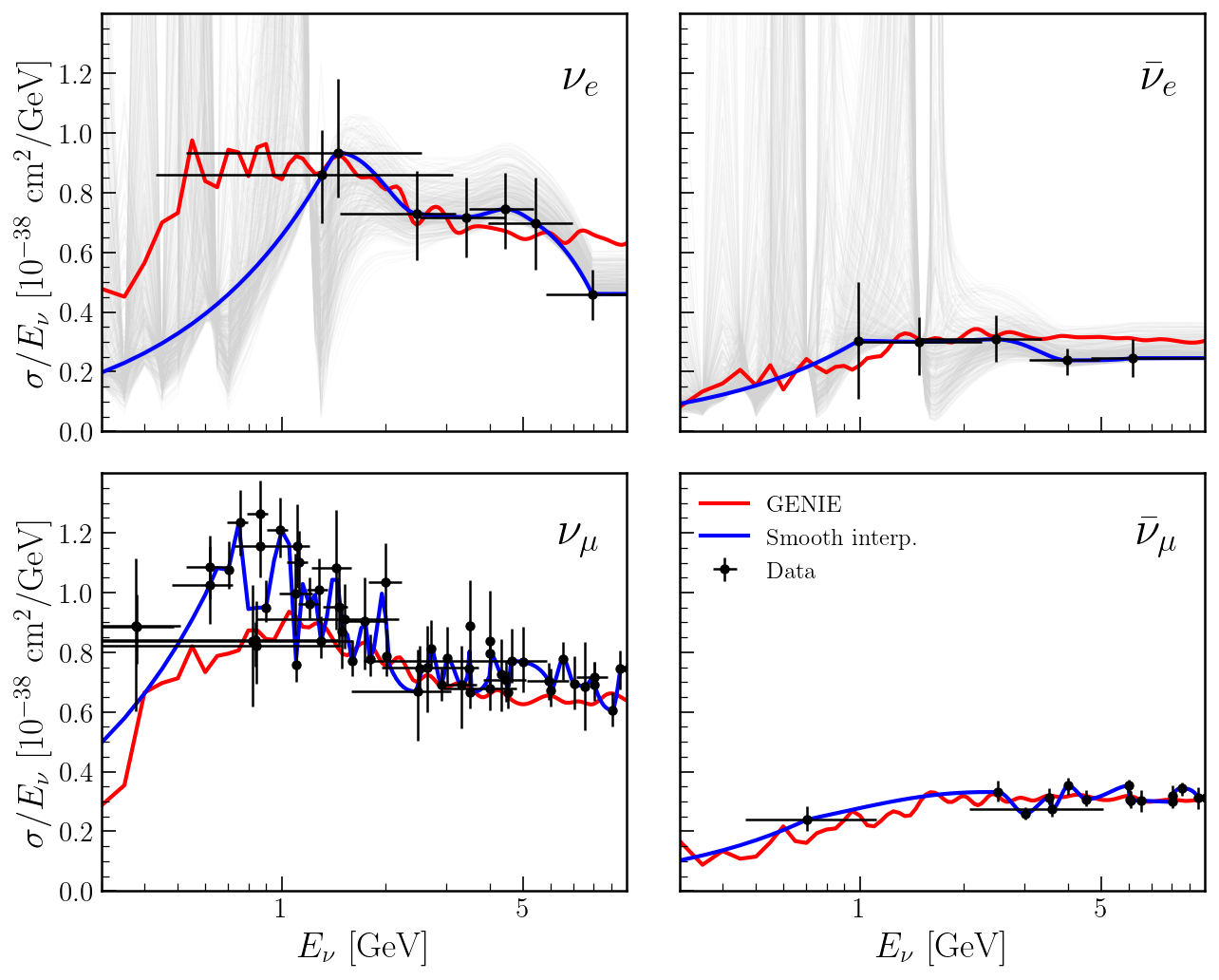}
\caption{
All panels show the neutrino cross section per nucleon divided by
neutrino energy, $\sigma / E_\nu$, as a function of $E_\nu$, for four
interaction channels: $\nu_e$ (upper left), $\bar{\nu}_e$ (upper
right), $\nu_\mu$ (lower left), and $\bar{\nu}_\mu$ (lower right). In
each panel, the black data points with error bars (statistical and
systematic uncertainties combined) represent measurements from
different experiments. The red curves show the cross sections from
\textsc{GENIE}. The blue curves represent the interpolated
data-driven cross sections obtained from fitting the experimental data
(see Sec.~\ref{sec:cross_section}). The grey curves show representative allowed deformations in the electron-flavor channels, $\nu_e$ and $\bar{\nu}_e$, which are the channels varied in our analysis.} 

\label{fig:cross_section_spline}
\end{figure}

Figure~\ref{fig:cross_section_spline} shows the data points in black with
error bars (statistical and systematic uncertainties combined),
representing measurements from different experiments for the four
interaction channels: $\nu_e$ (upper left), $\bar\nu_e$ (upper right),
$\nu_\mu$ (lower left), and $\bar\nu_\mu$ (lower right). In each panel,
the red curves show the cross sections from \textsc{GENIE}~v2.8.4~\cite{Andreopoulos:2015wxa,Tena-Vidal:2021rpu}, while the blue curves represent the interpolated
data-driven cross sections. The grey curves show representative
allowed cross-section deformations. While the $\nu_\mu$ and
$\bar\nu_\mu$ sectors are constrained by a substantial body of modern
measurements, the $\nu_e$ and $\bar\nu_e$ datasets remain sparse and
statistically limited, being largely dominated by historical
measurements from Gargamelle and a handful of recent low-statistics
results. As a consequence, the grey band in the electron-flavor panels
is broad: deviations of several percent in the cross-section shape---
including deformations that mimic the appearance probability---remain
fully compatible with existing data.

\subsection{Implementation of data-driven systematics}\label{sec:globes_implementation}

To quantify the impact of theoretical assumptions such as lepton
universality and nuclear modeling, we first establish a reference
nominal scenario in which these assumptions are taken to hold.

\subsubsection{Nominal configuration and standard systematics}\label{Sec:nominal}

In the nominal configuration, we simulate the T2HK and DUNE experiments using \texttt{GLoBES}~\cite{Huber:2004ka,Huber:2007ji}, adopting configurations consistent with their respective design reports.

For T2HK, we use the JHF flux files assuming a beam power of $1.2,\mathrm{MW}$ and a total exposure of $2.5~\nu + 7.5~\bar\nu$ years. The reconstructed energy window spans $0.4$--$1.2~\mathrm{GeV}$ in $20$ equal bins, with an energy resolution $\sigma(E) = 8.5\%\sqrt{E/\mathrm{GeV}}$. Signal efficiencies are $50.5\%$ for the $\nu_e/\bar\nu_e$ appearance channels and $90\%$ for the $\nu_\mu/\bar\nu_\mu$ disappearance channels. Backgrounds to the appearance signal include misidentified CC $\nu_\mu$ events, NC interactions, and intrinsic $\nu_e/\bar\nu_e$ beam contamination. Normalization uncertainties range from $1$--$2.5\%$ for signal channels up to $5\%$ for backgrounds, with an additional $3\%$ energy-tilt uncertainty applied to all channels.

For DUNE, we use the auxiliary files provided alongside Ref.~\cite{DUNE:2021cuw}, corresponding to an approximate TDR configuration. We assume an exposure of $6.5\nu + 6.5\bar\nu$ years with a beam power of $1.2~\mathrm{MW}$ ($1.1\times 10^{21}$ POT/yr), and a $40~\mathrm{kton}$ fiducial mass for the FD. The reconstructed energy range is $0.5$--$10~\mathrm{GeV}$ in $125~\mathrm{MeV}$ bins. Migration matrices and post-smearing efficiencies are taken from Ref.~\cite{DUNE:2021cuw}. Backgrounds include charge-misidentified CC events, NC interactions, $\nu_\tau$ CC from oscillations, and intrinsic $\nu_e/\bar\nu_e$ contamination. Normalization uncertainties range from $5\%$ to $20\%$, following the TDR prescription.

Following these assumptions, our setup reproduces the nominal $\dcp$ sensitivity reported in Refs.~\cite{DUNE:2021cuw,Hyper-Kamiokande:2018ofw}.

\subsubsection{Data driven systematics}

In order to account for systematics, standard
\texttt{GLoBES} uses nuisance parameters and the pull method, which can contribute
to an overall normalization or energy tilt such as the ones described in the nominal configuration. 
In previous literature, in order to account for energy distortions (shape uncertainties), 
in the context of the near detector, energy bin-to-bin uncorrelated nuisance parameters
have been explored~\cite{Coloma:2021uhq, Miranda:2018yym}. 
Here we require a qualitatively different treatment: instead of using
uncorrelated nuisance parameters, we consider a smooth distortion of the cross
section, as detailed in
Sec.~\ref{sec:distorsions}. We therefore need to evaluate $\chi^2_{\text{GLoBES}}(f_{\alpha})$, with
$(\alpha=\nu_e,\bar{\nu}_e)$, in the presence of such a distortion, which
requires non-trivial modifications of \texttt{GLoBES} which are detailed in Appendix~\ref{App:systematics_globes}. Before moving on, it is important to discuss a subtlety. At the far detector there will be two types of $\nu_e$ events: those coming from oscillations and those already present in the beam, corresponding to the intrinsic beam contamination. As discussed in Sec.~\ref{subsec:fd_nd}, the ND will measure $\Phi_{\nu_e}\sigma_{\nu_e}$ and, even though we have already discussed that this does not translate into a direct cross-section measurement, it gives a much better determination of the intrinsic $\nu_e$ contamination expected at the FD.\footnote{The extrapolation will not be perfect due to differences in acceptance between the ND and FD.} Therefore, we implicitly assume this component to be much better constrained and only allow for distortions in the appearance channel, where the theoretical cross-section uncertainty will play a major role. Including also distortions in the intrinsic $\nu_e$ component would largely neglect the information provided by the ND measurement and would therefore be too pessimistic. The use of the ND also reinforces our procedure of not including distortions in the muon sector. Current data already constrain this sector much more strongly, but future long-baseline experiments will in addition have ND measurements of the $\Phi_{\nu_\mu}\sigma_{\nu_\mu}$ convolution, providing a further constraint on the muon-neutrino contribution.

Once $\chi^2_{\text{GLoBES}}(f_e)$ is obtained from \texttt{GLoBES}, the penalization of each deformation $f_e$ by current data is introduced as one-dimensional priors,
\begin{equation}\label{eq:chi_1d}
\chi^2_{\alpha,1}(f_{\alpha}) \approx \frac{1}{N_{\alpha}}\chi^2_{\alpha}(f_{\alpha}) , 
\end{equation}
Here $N_\alpha$ is the number of experimental data points in the corresponding channel (see Eq.~\eqref{eq:chisq_xsec}). The factor $1/N_\alpha$ normalizes the cross-section $\chi^2$ to a goodness-of-fit measure per data point, such that a deformation producing discrepancies of roughly one experimental standard deviation per point gives a contribution of order one. This allows $\chi^2_{\alpha,1}$ to play the role of a continuous analogue of the usual pull method~\cite{Huber:2007ji,Fogli:2002pt}. We then marginalize over $f_\alpha$, yielding the following $\Delta\chi^2$:
\begin{equation}\label{eq:chi_marg} \Delta\chi^2 = \min_{f_\alpha}\left\{ \chi^2_{\text{GLoBES}}(f_{\alpha})+ \sum_{\alpha}\chi^2_{\alpha,1}(f_{\alpha}) \right\} \end{equation}

Intuitively, the deformations can vary during the minimization without receiving a significant penalty, as long as their shifts remain small compared to the width of their priors.

As already discussed in Sec.~\ref{sec:nuscope}, the experiment $\nu$SCOPE will provide two types of relevant measurements. It can directly measure $\sigma_{\nu_\mu}$ and $\sigma_{\bar{\nu}_\mu}$ via neutrino tagging, and it can indirectly constrain $\sigma_{\nu_e}$ and $\sigma_{\bar{\nu}_e}$ by measuring $R_{\text{NBOA}}$, defined in Eq.~\eqref{eq:RNBOA}, with an assumed precision of $2\%$. This information is incorporated as an additional Gaussian prior in Eq.~\eqref{eq:chi_marg}, of the form

\begin{equation}
\chi^2_{\nu\text{SCOPE}}(f_{\alpha})= \cfrac{\left(R_{\text{NBOA}}(f_\alpha)-R_{\text{NBOA}}\right)^2}{(0.02\, R_{\text{NBOA}})^2}
\end{equation}
where $R_{\text{NBOA}}$ is evaluated at zero distortion.\footnote{To compute $R_{\text{NBOA}}$, the virtual flux from Fig.~30 of Ref.~\cite{Acerbi:2025wzo} has been used.} Since, for simplicity, we consider distortions only in $\sigma_{\nu_e,\bar{\nu}_e}$ and do not introduce distortions in the muon sector, we do not impose additional priors beyond those associated with NBOA. $\nu$SCOPE will measure the muon-neutrino cross sections on a bin-by-bin basis using neutrino tagging, which allows to treat them as effectively known to the level of precision required for this study.

\section{The \dcppdf determination}
\label{sec:results}

We now present the main results of this work. Since the smooth cross-section deformations target the electron-flavor sector, the observable most affected is the CP-violation determination. These deformations could also impact the $\theta_{23}$ octant determination, but in a milder way; this is discussed separately in Appendix~\ref{App:theta23}.

We fix the oscillation parameters to the central values of the NuFIT~6.1
global fit~\cite{Esteban:2024eli} for normal ordering (NO), and account
for the corresponding uncertainties through marginalization. Results are
presented for NO throughout; we have verified that inverted ordering (IO)
leads to qualitatively similar conclusions. This is expected, as the
smooth deformations primarily affect the cross section, which is largely
independent of the mass ordering, while the differences in the appearance
probability between NO and IO are not large enough to significantly modify
the results. For each value of $\dcp^{\mathrm{true}}$, we compute the
significance at which CP conservation ($\dcp \in {0,\pi}$) can be rejected.
Since the significance vanishes at
$\dcp^{\rm true} = 0, \pm 180^\circ$, where the tested hypothesis
coincides with the true one, the periodicity of $\cos\dcp$ brings the
curve back to zero at both edges and the center of the range,
displaying the characteristic "McDonalds" shape.

\subsection{Sensitivities of DUNE and T2HK with Data-Driven Systematics}

\begin{figure}[t]
\centering
\includegraphics[width=0.48\linewidth]{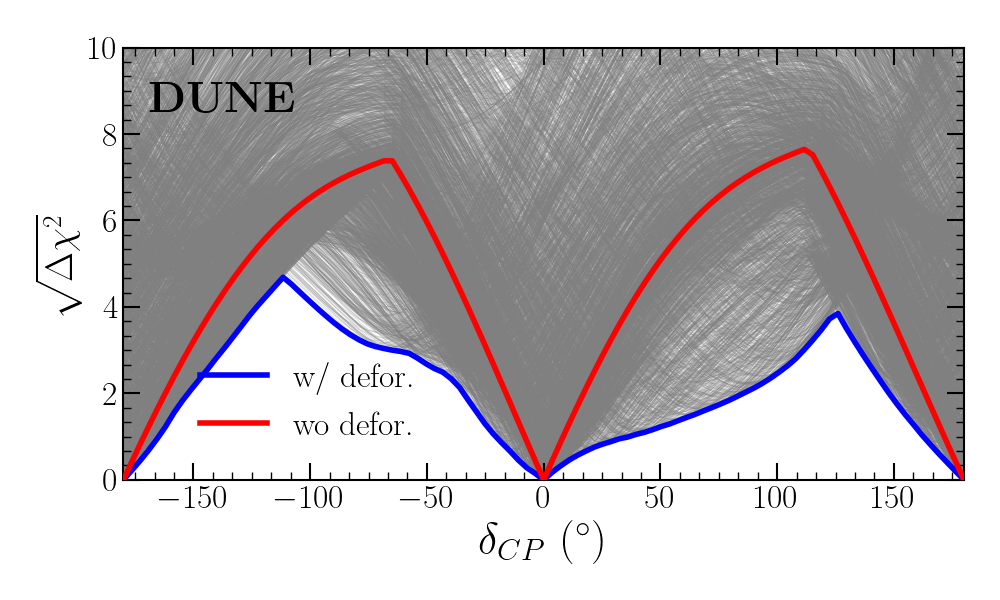}
\includegraphics[width=0.48\linewidth]{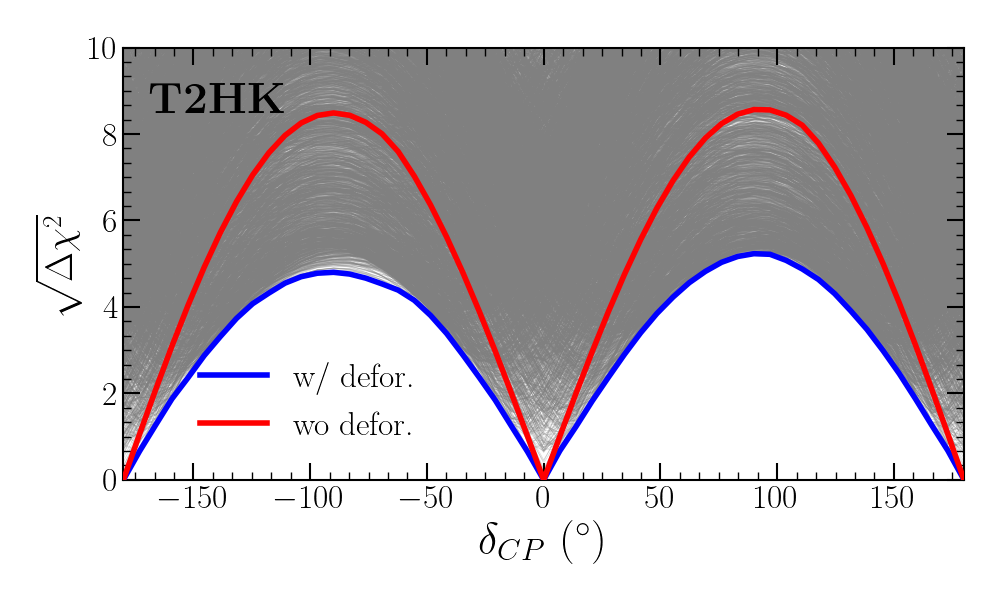}
\caption{CP-violation sensitivity expressed as $\sqrt{\Delta\chi^2}$ 
for rejecting CP conservation, as a function of $\dcp^{\mathrm{true}}$. 
\emph{Left panel:} DUNE, assuming $6.5\,\nu + 6.5\,\bar\nu$ years of 
running. 
\emph{Right panel:} T2HK, assuming $2.5\,\nu + 7.5\,\bar\nu$ years of 
running. 
The red curves show the nominal analyses without cross-section 
deformations. The grey curves correspond to individual smooth 
deformations $f_e(E)$ compatible with current cross-section data. 
The blue curves show the envelope, defined as the worst-case 
sensitivity obtained after marginalizing over all such deformations 
at each value of $\dcp^{\mathrm{true}}$. All other oscillation 
parameters are marginalized over.}
\label{fig:result_splines}
\end{figure}

We begin by quantifying how much of the CP-violation sensitivity relies on
the assumptions of lepton universality and accurate nuclear modelling.
Figure~\ref{fig:result_splines} shows the sensitivity as a function of
$\dcp^{\mathrm{true}}$, assuming $6.5~\nu + 6.5~\bar\nu$ years for DUNE
(left panel) and $2.5~\nu + 7.5~\bar\nu$ years for T2HK (right panel). In
both panels, the red curves correspond to the nominal analyses described
in Sec.~\ref{Sec:nominal}, reaching $\sqrt{\Delta\chi^2} \sim 7$ for DUNE
and $\sim 8.6$ for T2HK at maximal CP violation ($\dcp^{\mathrm{true}} \sim \pm 90^\circ$). At DUNE, 
the wide-band beam covers the region around the first and second
oscillation maxima and part of the sensitivity comes from the spectral
shape through the $\cos\Delta\,\cos\dcp$ term, rather than from the
overall rate asymmetry alone. In addition, the matter potential is
sizable at $L = 1300$~km and enters with opposite sign for neutrinos
and antineutrinos, explicitly breaking the $\dcp \to -\dcp$ symmetry.
As a result the DUNE sensitivity curve is slightly asymmetric, with the two lobes
differing in height and shape. 
At T2HK, instead,
the narrow-band beam sits at the first oscillation maximum, where
$\Delta \simeq \pi/2$ and the CP-odd term of Eq.~\eqref{eq:Pmat}
reduces to $-\sin\Delta\,\sin\dcp$. Therefore, the sensitivity is 
dominated by the neutrino--antineutrino rate asymmetry and the profile
essentially tracks $|\sin\dcp|$, giving two smooth, rounded lobes.
Moreover, matter effects are small at $L = 295$~km, so the
approximate $\dcp \to -\dcp$ symmetry of the vacuum probability is
preserved and the two lobes have nearly equal height.

The grey curves show the effect of individual smooth cross-section deformations $f_e(E)$ compatible with existing data (see Sec.~\ref{sec:distorsions}), while the blue curve represents the envelope of these solutions, defined as the worst-case deformation at each $\dcp^{\mathrm{true}}$.  This provides a conservative, data-driven estimate of the sensitivity. After marginalizing over these deformations, the sensitivity is reduced to $\sqrt{\Delta\chi^2} \sim 4$ for DUNE and $\sim 4.7$ for T2HK. Note that, although the T2HK flux peaks around $0.6$ GeV, below the central energies of the lowest-energy cross-section measurements shown in Fig.~\ref{fig:cross_section_spline}, these measurements are obtained with broad, non-monochromatic neutrino fluxes, resulting in large energy uncertainties and therefore constraining the cross section over a wider energy range, including the sub-GeV region.

Both experiments exhibit a similar relative degradation, corresponding to a reduction of roughly $40$–$50\%$ in significance. This indicates that a substantial fraction of the CP sensitivity is controlled by assumptions on the $\nu_e/\nu_\mu$ cross-section ratio, making the $\sigma_{\nu_e}/\sigma_{\nu_\mu}$–$\dcp$ degeneracy an important limitation at this level of precision.

\subsubsection{Impact of \nupdf SCOPE}

\begin{figure}[t]
\centering
\includegraphics[width=0.48\linewidth]{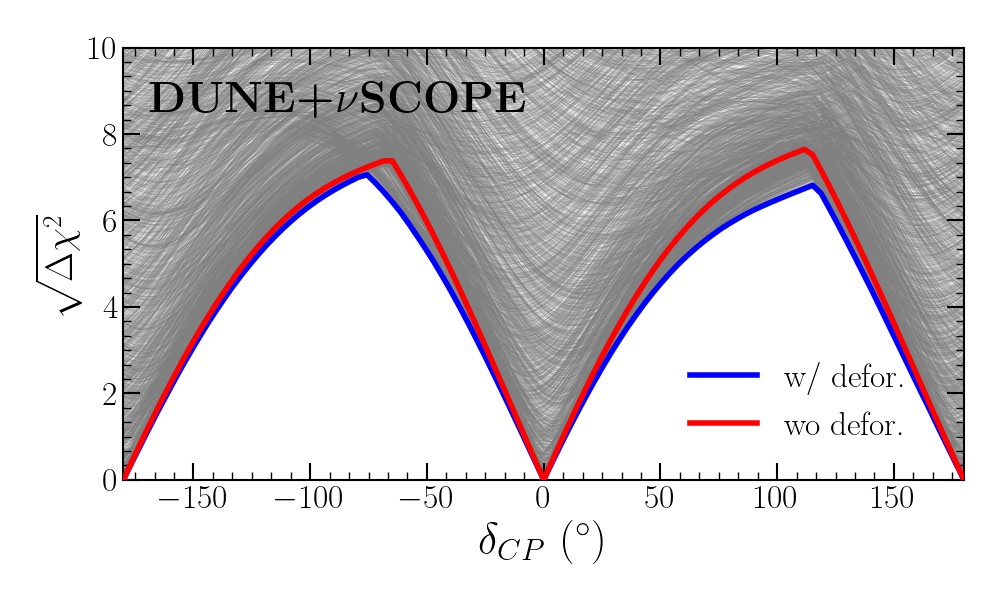}
\includegraphics[width=0.48\linewidth]{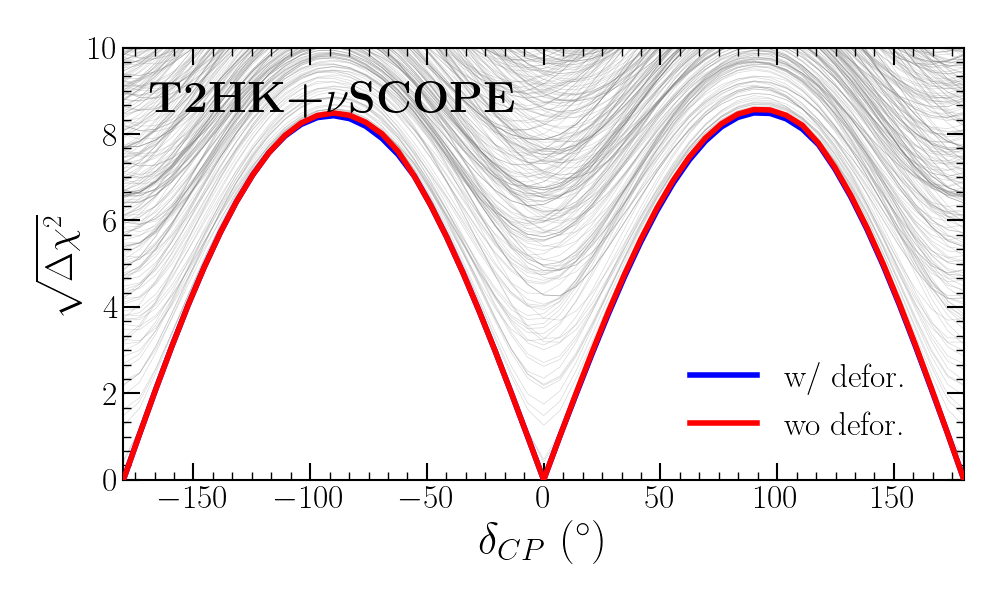}
\caption{Same as Fig.~\ref{fig:result_splines}, but including the 
prospective $\nu$SCOPE prior on $R_{\mathrm{NBOA}}$ at $2\%$ precision. 
\emph{Left panel:} DUNE\,+\,$\nu$SCOPE. 
\emph{Right panel:} T2HK\,+\,$\nu$SCOPE. 
The red curves show the nominal analyses without cross-section 
deformations. The grey curves correspond to smooth deformations 
compatible with both current cross-section data and the $\nu$SCOPE 
constraint. The blue curves show the worst-case envelope obtained 
after marginalizing over all such deformations. All other oscillation 
parameters are marginalized over.}
\label{fig:result_splines_final}
\end{figure}

Figure~\ref{fig:result_splines_final} shows the same analysis after incorporating the prospective $\nu$SCOPE measurement of $R_{\mathrm{NBOA}}$ as an external prior, as described in Sec.~\ref{sec:globes_implementation}. The grey curves now represent only those deformations that remain allowed after imposing the $2\%$ constraint from $\nu$SCOPE, and their spread is visibly reduced. As a result, the sensitivity is largely restored in both experiments: the blue envelope at maximal CP violation reaches $\sqrt{\Delta\chi^2} \sim 6.5$ for DUNE (left panel) and $\sim 8.6$ for T2HK (right panel), recovering most of the sensitivity lost in the standalone analysis.

This improvement demonstrates that, although $\nu$SCOPE constrains $\sigma_{\nu_e}$ only indirectly through the flux-weighted ratio $R_{\mathrm{NBOA}}$, its $\sim 2\%$ precision is sufficient to eliminate the most dangerous smooth deformations and bring the sensitivities close to their nominal reach.

\section{Summary and conclusions}\label{sec:summary_conclusions}

Determining whether CP is violated in the leptonic 
sector remains one of the major open questions in 
neutrino oscillation physics, and future LBL 
experiments such as DUNE and T2HK are expected to 
provide a decisive answer. As the field enters the precision era, 
however, the ultimate reach of these 
experiments is no longer limited by statistics alone, but 
by the control of systematic uncertainties. In particular, the 
determination of $\dcp$ relies on 
the near-to-far extrapolation and, crucially, on the assumed knowledge 
of the cross-section ratios $\sigma_{\nu_e}/\sigma_{\nu_\mu}$ and 
$\sigma_{\bar\nu_e}/\sigma_{\bar\nu_\mu}$, which are commonly 
treated as known at the few-percent level under the assumptions 
of lepton universality and 
accurate nuclear modeling. Mismodeling of nuclear physics, final-state interactions, 
or BSM effects that violate lepton universality can 
introduce unexpected energy-dependent 
distortions that affect $\nu_e$ and $\nu_\mu$ differently.

In this work, we have performed an agnostic, data-driven stress test of these assumptions by introducing smooth deformations of $\sigma_{\nu_e}$ and $\sigma_{\bar\nu_e}$ that can reproduce the $\dcp$-dependent spectral modulation in the appearance channel, while remaining consistent with existing cross-section data. 

Our main results can be summarized as follows:

\begin{itemize}
\item In the nominal configurations ($6.5~\nu + 6.5~\bar\nu$ years for DUNE and $2.5~\nu + 7.5~\bar\nu$ years for T2HK), the CP-violation sensitivity at $\dcp^{\mathrm{true}} = \pm 90^\circ$ reaches $\sqrt{\Delta\chi^2} \sim 7$ (DUNE) and $\sim 8.6$ (T2HK).

\item After marginalizing over the full set of allowed deformations, the sensitivity is reduced to $\sqrt{\Delta\chi^2} \sim 4$ for DUNE and $\sim 4.7$ for T2HK. This corresponds to a comparable relative degradation of $\sim 40$–$50\%$ in both experiments, showing that a substantial fraction of the CP sensitivity is driven by assumptions on the $\nu_e/\nu_\mu$ cross-section ratio. This establishes the $\sigma_{\nu_e}/\sigma_{\nu_\mu}$–$\dcp$ degeneracy as an important limitation at this level of precision.

\item Including the projected $\nu$SCOPE constraint on $R_{\mathrm{NBOA}}$ at the $\sim 2\%$ level restores the sensitivity to $\sqrt{\Delta\chi^2} \sim 6.5$ (DUNE) and $\sim 8.6$ (T2HK), effectively recovering the nominal performance in both cases.

\item The impact on the $\theta_{23}$ octant determination is significantly milder: the sensitivity is moderately reduced in the presence of cross-section deformations and largely recovered once the $\nu$SCOPE prior is included (see Appendix~\ref{App:theta23}).
\end{itemize}

Our results demonstrate that precise measurements of 
neutrino cross sections — such as those provided by a 
facility like $\nu$SCOPE — may play a crucial role for establishing leptonic 
CP violation in a robust and assumption-independent way.

\paragraph{Acknowledgments.} 

We thank Jacobo~López-Pavón and Pedro Machado for useful discussions.
JPP is supported by the National Natural Science Foundation of China (12425506 and 12375101). This work is also supported by State Key Laboratory of Dark Matter Physics.

\appendix

\section{Explicit construction of the deformations for our \texorpdfstring{$\delta_{\text{CP}}$}{deltaCP} analysis}
\label{App:deformation_details}

In this appendix, we explain how we construct the cross-section
deformations and assess their compatibility with existing data. For each channel $\alpha = \nu_e,\bar{\nu}_e$, we define a reference cross section
$(\sigma_\alpha/E)^{\rm interp}$ by applying a piecewise-cubic Hermite
interpolant (PCHIP)~\cite{Fritsch1984AMF} to the experimental points discussed in
Sec.~\ref{sec:cross_section_status}, over the energy range
$E\in[0.5,10]$~GeV, together with the physical condition
$\sigma(E\to0)=0$. This curve provides a smooth central value around
which we define the deformations. Our results depend little on its
precise shape: repeating the analysis with the GENIE prediction as the
reference cross section gives negligible changes.

We write the deformed cross section as
\begin{equation}
\left(\frac{\sigma_{\alpha}}{E}\right)^{\mathrm{def}}(E,f_{\alpha})
=
\left[1+f_{\alpha}(E)\right]
\left(\frac{\sigma_{\alpha}}{E}\right)^{\mathrm{interp}}.
\end{equation}
Although $f_\alpha(E)$ could in principle be arbitrary, we focus on
deformations that can mimic the spectral effect of changing
$\delta_{\rm CP}$. In particular, for Fig.~\ref{fig:result_splines},
we ask whether a CP-conserving scenario could be interpreted as
CP violating if the electron-flavor cross sections differ from the
reference curves by an amount still allowed by present data.

With this in mind, we choose
\begin{equation}
1+f_\alpha(E;\delta)
=
1+a_{\alpha}\Bigg(
\frac{P_{\mu\alpha}(E;\delta_{\rm ref}=0,180^{\circ})}
     {P_{\mu\alpha}(E;\delta)} -1\Bigg),
\qquad
\alpha=\nue,\anue,
\label{eq:def_shape}
\end{equation}
The parameters $\delta$ and $a_\alpha$ are sampled independently from uniform distributions over $\delta\in[-180^\circ,180^\circ]$ and $a_\alpha\in[-1,1]$, respectively. To avoid unphysical deformations, we further require that $1+f_\alpha(E;\delta)>0$. The ratio of oscillation
probabilities gives the energy dependence needed for the cross-section
deformation to compensate for a change in $\delta_{\rm CP}$, up to the
overall factor $a_\alpha$.  This choice of deformation is sufficient for our purpose,
since these are precisely the deformations with the greatest potential to
degrade the $\delta_{\rm CP}$ sensitivity. A more general set of deformations
would also contain these directions, but scanning the full functional space
would be numerically inefficient.

Several cross-section measurements cover broad energy intervals.
To compare with these measurements, we average the deformed cross
section over each experimental bin:
\begin{equation}
\left(\frac{\sigma_\alpha}{E}\right)^{\mathrm{avg}}_i(f_\alpha)
=
\frac{1}{\Delta E_i}
\int_{E_i^{\mathrm{min}}}^{E_i^{\mathrm{max}}}
dE\,
\left(\frac{\sigma_\alpha}{E}\right)^{\mathrm{def}}(E,f_\alpha),
\end{equation}
where
$\Delta E_i=E_i^{\mathrm{max}}-E_i^{\mathrm{min}}$ is the width of
the $i$-th bin.

For each flavor channel, we quantify the agreement with the data through
\begin{equation}
\chi^2_{\alpha}(f_{\alpha})
=
\sum_{i=1}^{N}
\left[
\frac{
(\sigma_\alpha/E)_i^{\mathrm{data}}
-
(\sigma_\alpha/E)_i^{\mathrm{avg}}(f_\alpha)
}{
\delta\sigma_i
}
\right]^2,
\end{equation}
where $\delta\sigma_i$ is the experimental uncertainty on the measured
$(\sigma_\alpha/E)_i^{\mathrm{data}}$, and $N$ is the number of available
data points in that channel. For a fixed deformation and independent
Gaussian uncertainties, this statistic approximately follows a
$\chi^2$ distribution with $N$ degrees of freedom.

We generate $10^5$ deformations and evaluate their agreement with the
cross-section data using Eq.~\eqref{eq:chisq_xsec}. These deformations
are shown in grey in Fig.~\ref{fig:cross_section_spline} and are used
throughout Sec.~\ref{sec:results}. Note that none of the deformations is
directly rejected. Instead, in the oscillation analysis, each deformation
is included with a penalty given by the normalized quantity defined in
Eq.~\eqref{eq:chi_1d}. Dividing by the number of data points gives a
penalty of order one for a deformation whose deviations from the
measurements are typically of order one experimental standard
deviation. Deformations that depart strongly from the data receive
correspondingly large penalties and therefore cannot significantly lower
the $\Delta\chi^2$ value of Eq.~\eqref{eq:chi_marg}. A similar strategy is used for the $\theta_{23}$ analysis shown in
Appendix~\ref{App:theta23}.

\section{Implementation of data-driven systematics in GLoBES}
\label{App:systematics_globes}

The main goal of this appendix is to describe how we compute $\chi^2_{\text{GLoBES}}(f_e)$. As shown in Eq.~\eqref{eq:NFD_spectrum}, the event rate depends on the convolution of the flux, cross section, and oscillation probability. In \texttt{GLoBES}, fluxes and cross sections are provided as inputs, while probabilities are computed internally. Since the observable depends on their convolution, deformations of the cross section can be effectively implemented by rescaling the probability.

For a given deformation $f_e(E)$, this is implemented by rescaling the appearance probability,
\begin{equation}
P_{\mu e}(E) \;\rightarrow\; P_{\mu e}(E)\,\left[1 + f_e(E)\right]\,,
\end{equation}
which is applied only to channels involving $\nu_e$ charged-current interactions, leaving all other channels unchanged.

In the standard version of \texttt{GLoBES}, this cannot be implemented directly, since probabilities are computed independently of the flux and cross section. As a result, a naive rescaling of the probability would affect all channels, rather than only those involving $\sigma^{\text{CC}}_{\nu_e}$. This therefore requires a version in which detection effects can be handled at the probability level.

We use the extended version of \texttt{GLoBES} developed in Ref.~\cite{Kopp:2025ffx}, where probabilities are evaluated channel by channel with access to the corresponding flux and cross section, here in the context of including production and detection effects induced by EFTs. We adapt this implementation to include the energy-dependent deformation $f_e(E)$. For each deformation, we compute $\chi^2_{\text{GLoBES}}(f_e)$. The results are then combined with the prior of Eq.~\eqref{eq:chi_marg}, and the marginalization over $f_e$ is performed externally.

\section{Determination of the \tzpdf octant}
\label{App:theta23}

Although the main focus of this work is the determination of $\delta_{\text{CP}}$, for completeness we study here the impact of relaxing the assumptions of lepton universality and nuclear modeling on the determination of the $\theta_{23}$ octant. The measurement of the $\theta_{23}$ octant is intrinsically challenging: around maximal mixing ($\theta_{23} \simeq 45^\circ$), the oscillation probabilities depend only weakly on the octant, since the leading terms are approximately symmetric under $\theta_{23} \to 90^\circ - \theta_{23}$. As a result, the sensitivity arises only from subleading effects, primarily through the appearance channel, and is therefore more limited than that of the CP-violation measurement.

\begin{figure}[h!]
\centering
\includegraphics[width=\linewidth]{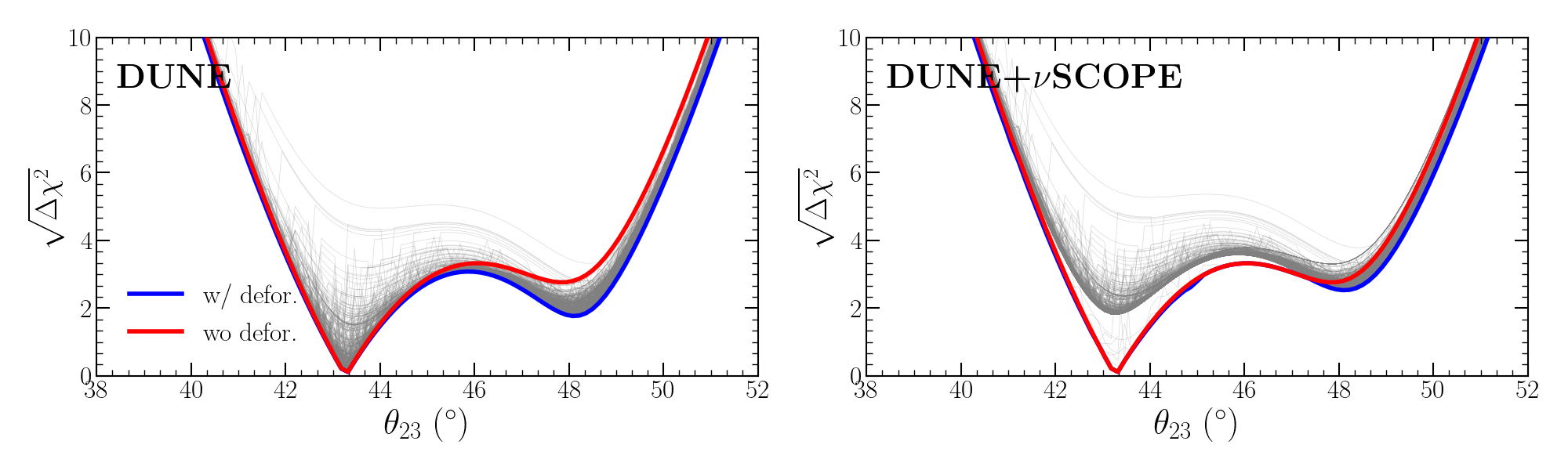}
\caption{Impact of data-driven cross-section systematics on the
determination of the $\theta_{23}$ octant at DUNE, assuming
$\theta_{23}^{\mathrm{true}} = 43.3^\circ$ (first octant). The
sensitivity to reject the wrong octant is shown as
$\sqrt{\Delta\chi^2}$ as a function of
$\theta_{23}^{\mathrm{true}}$. In both panels, the red curve
corresponds to the nominal analysis without cross-section
deformations, the grey curves show individual allowed deformations,
and the blue curve is the result after marginalizing over all
deformations. \emph{Left panel:} DUNE standalone. \emph{Right panel:}
DUNE combined with the prospective $\nu$SCOPE prior. All other
oscillation parameters are marginalized over.}
\label{fig:result_splines_23}
\end{figure}

\begin{figure}[h!]
\centering
\includegraphics[width=\linewidth]{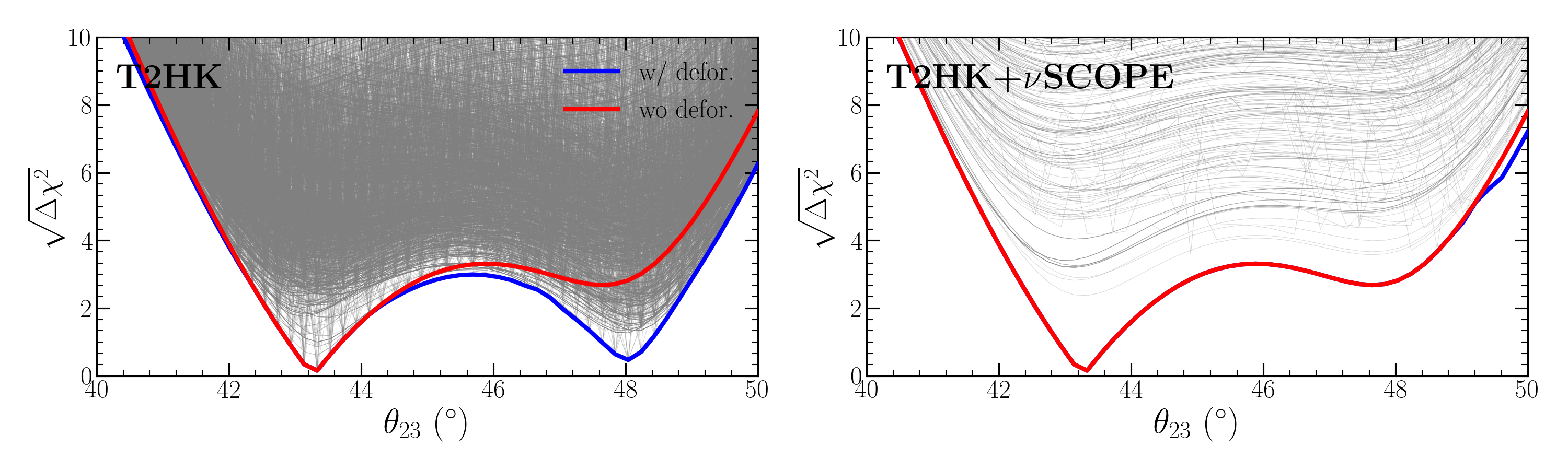}
\caption{Impact of data-driven cross-section systematics on the
determination of the $\theta_{23}$ octant at T2HK, assuming
$\theta_{23}^{\mathrm{true}} = 43.3^\circ$ (first octant). The
sensitivity to reject the wrong octant is shown as
$\sqrt{\Delta\chi^2}$ as a function of
$\theta_{23}^{\mathrm{true}}$. In both panels, the red curve
corresponds to the nominal analysis without cross-section
deformations, the grey curves show individual allowed deformations,
and the blue curve is the result after marginalizing over all
deformations. \emph{Left panel:} T2HK standalone. \emph{Right panel:}
T2HK combined with the prospective $\nu$SCOPE prior. All other
oscillation parameters are marginalized over.}
\label{fig:result_splines_23_T2HK}
\end{figure}

We present our results in Fig.~\ref{fig:result_splines_23} and Fig.~\ref{fig:result_splines_23_T2HK}. Assuming a true value of $\theta_{23} = 43.3^\circ$ (first octant), corresponding to the global-fit value~\cite{Esteban:2024eli}, we evaluate the significance at which the second octant can be rejected after marginalizing over all oscillation parameters. In the nominal analysis (i.e., without cross-section deformations), DUNE achieves a rejection of the wrong octant at the $\sim2.6\sigma$ level, while T2HK reaches a sensitivity of $\sim3\sigma$, already significantly lower than for $\delta_{\text{CP}}$. After marginalizing over the full set of allowed deformations, the sensitivity decreases to $\sim1.7\sigma$ for DUNE and to $\sim0.3\sigma$ for T2HK, corresponding to degradations of about $1\sigma$ and $3\sigma$, respectively. Although sizable, this reduction is significantly milder than the $\sim 3$--$4\sigma$ loss observed in the CP-violation determination. This difference can be understood as follows: while the CP-violating signal relies on a subleading interference term that can be efficiently mimicked by smooth energy-dependent distortions, the octant sensitivity is already limited and is driven mainly by an overall modulation of the appearance rate, so the set of deformations that can further degrade it is more restricted. Including the prospective $\nu$SCOPE measurement as an external prior largely restores the sensitivity to the $\theta_{23}$ octant, bringing it back to $\sim 2.4\sigma$ for DUNE and $\sim 3\sigma$ for T2HK.

\bibliographystyle{JHEP} 
\bibliography{Refs}

\end{document}